\documentclass[10pt,journal,letterpaper]{IEEEtran}

\usepackage{amssymb}
\usepackage{graphicx}
\usepackage{color}
\usepackage{amsmath}
\newcommand{\SINR}{{\rm SINR}}

\newcommand{\BR}{{\rm BR}}
\newcommand{\atana}[1]{\arctan_{\alpha} \left( {#1}  \right) }
\newcommand{\argmax}{\operatornamewithlimits{argmax}}

\newtheorem{definition}{Definition}[section] 
\newtheorem{lma}{Lemma}[section]
\newtheorem{proposition}{Proposition}[section]
\newtheorem{cor}{Corollary}[section]
\newtheorem{thm}{Theorem}[section]

\newtheorem{remark}{Remark}[section] 

\begin{document}
\title{Spatial SINR Games of Base Station Placement and Mobile Association}

\author{Eitan~Altman,~\IEEEmembership{Fellow,~IEEE,}
Anurag~Kumar,~\IEEEmembership{Fellow,~IEEE,} 
Chandramani~Singh,~\IEEEmembership{Student Member,~IEEE,}\\
and~Rajesh~Sundaresan,~\IEEEmembership{Senior Member,~IEEE}
\thanks{This is an extended version of a paper that appeared in IEEE Infocom 2009.}
\thanks{This work was supported by an INRIA Associates program DAWN, and also by the Indo-French Centre for the Promotion of Advanced Research (IFCPAR), Project No.~4000-IT-A.}
\thanks{Eitan~Altman is with INRIA, Sophia-Antipolis, France~(email: Eitan.Altman@sophia.inria.fr). Anurag~Kumar, Chandramani~Singh and Rajesh~Sundaresan are with the Department of Electrical Communication Engineering
Indian Institute of Science Bangalore, India~(email: \{anurag,~chandra,~rajeshs\}@ece.iisc.ernet.in).}
}

\maketitle


\begin{abstract}
We study the question of determining locations of base stations that may belong to the same or to competing service providers. We take into account the impact of these decisions on the behavior of intelligent mobile terminals who can connect to the base station that offers the best utility. The signal to interference and noise ratio is used as the quantity that determines the association. We first study the SINR association-game: we determine the cells corresponding to each base stations, i.e., the locations at which mobile terminals prefer to connect to a given base station than to others. We make some surprising observations: (i) displacing a base station a little in one direction may result in a displacement of the boundary of the corresponding cell to the opposite direction; (ii) A cell corresponding to a BS may be the union of disconnected sub-cells. We then study the hierarchical equilibrium in the combined BS location and mobile association problem: we determine where to locate the BSs so as to maximize the revenues obtained at the induced SINR mobile association game. We consider the cases of single frequency band and two frequency bands of operation. Finally, we also consider hierarchical equilibria in two frequency systems with successive interference cancellation.

\end{abstract}

\section{Introduction}
In this paper we study some hierarchical decision making problems arising in the uplinks of cellular networks. We first address the problem of association: given multiple base stations~(BS) capable of providing services to a mobile located at a given point in the region of operation, to which BS should the mobile connect? This is studied in a non-cooperative context where each mobile connects to the BS that provides it with the best signal to interference and noise ratio~(SINR). The associations determine the {\em cells} corresponding to each BS. We characterize the nature of cells as a function of BS locations. 

We then consider the problem of determining the locations of base stations, taking into account the behavior of the mobiles that will be induced by the location decisions. We study cases where the BSs cooperate (e.g., they belong to the same service provider) and those where they compete with each other. The latter scenario results in a location game between the BSs.

\paragraph*{Related work}
Plastria \cite{plastria} presented an overview of research on locating one or more new facilities in an environment where competing facilities already exist. Gabszewicz and Thisse \cite{GaThi} provided another general overview on location games. 
Mazalov and Sakaguchi \cite{mazalov} and references therein studied competition over prices of goods between facilities that have fixed positions. They then derived the equilibrium allocation of customers. Such games, as well as hierarchical games in which firms compete for location or over prices which then determine the customer-allocation equilibrium, were introduced by Hotelling \cite{hotelling} in 1929. When considering such games over a finite line segment with two firms, the models under appropriate conditions give rise to a partition of the segment into two convex subsegments or ``cells'' as introduced in our context.

An interesting difference between the settings above and our setting, which is also defined on a finite line segment, is that in our case more complex cells are obtained at equilibrium. This is due to the difference in the cost structure in the cellular context. Hotelling \cite{hotelling} considered a general cost related to the distance between the customer and the firm it chooses; this cost however depended only on the distance and not on the actual location of the firm. This does not hold in our case: the throughput of the mobile depends on the interference at the base station which in turn depends on the location of the base station. We finally note that in our model, the power of a mobile, which can be considered as the ``cost'', is fixed, while it attempts to maximize the utility, i.e., throughput.   

Aram et al.~\cite{ASSK} study coalition based joint resource procurement and resource allocation in wireless networks using the framework of cooperative game theory. They consider a set of operators and customers with predetermined customer-operator associations. All the operators together place base stations, procure spectrum, and allocate channels to the common pool of customers.  Doing so is shown to be optimal even when the operators are selfish. However, it is assumed that operators can divide the aggregate earned utility in any arbitrary way. In another work, Aram et al.~\cite{ASS} extend the analysis to nontransferable utilities.  

After the appearance of our initial work~\cite{AKSS}, Ramanath et~al.~\cite{RAKKT} have studied the joint placement of two base stations that use same frequency. The utilities of BSs come from the notion of $\alpha$-fairness. They also consider a multicell scenario where each cell has one BS, and all BSs operate on the same frequency. The goal is to optimally place the BSs in their respective cells so as to maximize their $\alpha$-fair utilities. Silva et~al.~\cite{SADTJ} consider an association problem where mobiles have hard average throughput constraints, and the objective of optimal association is to minimize the aggregate power consumption of all the mobiles in the network. They also study the downlink scenario under the assumption that neighboring BSs operate in orthogonal channels. 
Again the objective is to find an optimal association that minimizes the aggregate power consumption of the BSs. Kasbekar et~al.~\cite{KAS} consider a joint problem of mobiles' association and charging and spectrum leasing by service providers. 

A description of the model studied in this paper and the notation used can be found in Section \ref{sec:model} and Appendix \ref{app:prop-fluid}.

\paragraph*{Our contributions}
First, we consider a code division multiple access~(CDMA) system where BSs perform single mobile decoding. We derive analytical expressions for the cell boundaries in the case where BSs are on the same frequency band~(Section \ref{subsec:sinrEqAssoc}). This allows us to study the geometric properties of cells as a function of the locations of the BSs. We then study the hierarchical equilibrium in the combined BS location and mobile association problem, i.e., we determine where to locate the BSs so as to maximize the revenues obtained on the induced SINR-based mobile association game~(Section \ref{subsec:hierarchical}).  We also do the analogous analyses for the case where BSs are on different frequency bands~(Section \ref{sec:twoFrequencies}). Subsequently, we consider BSs capable of successive interference cancellation~(SIC) decoding. After a discussion on the association problem and the single frequency band case~(Section \ref{sec:sicsinglefreq}), we analyze the case of different frequency bands and give a complete characterization of the resulting equilibria~(Section \ref{sec:sictwofreq}).

While the main body of our work assumes mobiles placed over the line-segment $[-L,L] \subset \mathbb{R}$, we discuss the extension to two dimensional deployments in Appendix~\ref{app:2D}.

\section{The model and notation}
\label{sec:model}
Our focus is on communication in the uplink direction, i.e., from mobiles to BSs. Mobiles and BSs lie in the two-dimensional space $\mathbb{R}^2$. A large number of mobiles are placed uniformly over the segment $[-L,L]$ on the first of the coordinate axes. The fluid approximation is obtained for the infinitely large population of mobiles. For details see Appendix~\ref{app:prop-fluid}. There are two BSs, BS~1 and BS~2, located at $(x_1,1)$ and $(x_2,1)$, respectively~(say on the top of a flat building whose height is one unit). BSs cooperate if they belong to the same operator, and compete if they belong to different operators. We allow placements of BSs outside the area where mobiles exist, i.e., $x_j < -L$ and $x_j > L$ are allowed for $j = 1,2$. In the following, we use only the first coordinates to specify locations (with the understanding that second coordinates are 0 in case of mobiles, and 1 in case of BSs).

Transmitters are point sources radiating in two-dimensional space with circular wavefronts (respectively, three-dimensional with spherical wavefronts). We consider a power law path loss model with exponent $\alpha$; i.e., the power from a radio transmitter attenuates as distance raised to the power $\alpha$ (see Appendix \ref{app:prop-fluid}). A mobile located at $y$ has a channel ``gain'' of $[(y-x_j)^2+1]^{-\alpha/2}$ to BS~$j$. All mobiles are assumed to transmit at a power such that the power density along the line is unit power per unit length.
Thus the total transmitted power is $2L$.
Thermal noise at the BSs is assumed to be Gaussian with noise variance $\sigma^2$ per sample.

At any time, each mobile is \emph{associated} with exactly one BS. Let $A_j \subset [-L,L]$ be such that the mobiles in $A_j$ are associated with BS~$j$. $A_j$ will be called {\em cell} $j$. The utility of a mobile at $y$ is assumed to be a nondecreasing function of the {\em SINR density} at $y$, as seen at the BS to which the mobile is associated. The SINR density depends on the interference model under consideration, which we discuss next.  

\subsection{Interference models}
Mobiles that connect to a particular BS may or may not cause interference to the other BS depending on whether the BSs operate on the same or different radio frequency~(RF) bands. We consider both the cases in this paper. The case in which the same frequency band (channel) is used at both the BSs occurs if the wireless network operates in an unlicensed band; in such a case, BSs belonging to different networks (or providers) may use the same RF band.  We call this \emph{the single-frequency case}. If the wireless network operates in licensed RF bands, two neighboring BSs would operate in disjoint RF bands. We call this \emph{the two-frequencies case}. We now discuss the useful power collected and interference seen at a BS in the single- and two-frequencies cases. For this purpose, it is useful to define the following functions. Define
\begin{equation}
  \label{eqn:g}
  g(y) := [1+y^2]^{-\alpha/2}.
\end{equation}
For a set $S \subseteq [-L,L]$ and candidate BS location $x$, define
\begin{equation}
  \label{eqn:E-x-S}
  E(x,S) := \int_S g(y-x) ~dy.
\end{equation}
and $E^o(x) := E(x,[-L,L])$. The dependence of $g$, $E$, and $E^o$ on $\alpha$ is understood.

In the following we consider a CDMA system where BSs perform single mobile decoding, i.e., while decoding any mobile's signal, they treat all other mobiles' received signals as interference. Subsequent descriptions of SINR-equilibrium and hierarchical equilibrium are also for such a system. Analogous notions for SIC decoding are defined in Sections~\ref{sec:sicsinglefreq}~(single-frequency case) and~\ref{sec:sictwofreq}~(two-frequencies case). 
   
\subsubsection{The single-frequency case} In this case, power from all the mobiles is received at both the BSs. The total received power at BS~$j$ located at $x_j$ is therefore given by $E(x_j, [-L,L]) = E^o(x_j)$. All of this received power will clearly be interference to a mobile at $y$ because the mobile's own contribution to this is infinitesimal.

With this interference interpretation for $E^o(x_j)$, we now highlight some of its properties. It is straightforward to see via change of variables that
\begin{equation}
\label{Int3}
  E^o(x)  = \int_{-L-x}^{L-x} g(y) ~dy =  \int_{\arctan(-L-x)}^{\arctan(L-x)} (\cos \theta)^{\alpha-2} ~d\theta. 
\end{equation}
Closed form expressions are available for $E^o$ when $\alpha$ takes integer values. In particular, for $\alpha=2$ we get
\begin{equation}
\label{Int4}
E^o(x) =
    \arctan(L-x) + \arctan(L+x),
\end{equation}
and for $ \alpha=1$ we get
\begin{equation}
E^o(x) =
   \mbox{arcsinh}(L - x) + \mbox{arcsinh}(L + x).
\label{Int5}
\end{equation}
The above expressions motivate the following definition of the $\alpha$-parametric function
\[
  \atana{x} := \int_0^x g(y) ~dy, ~x \in \mathbb{R}.
\]
Then clearly $\atana{\cdot}$ is an odd function\footnote{It is an odd function because $\atana{-x} = - \atana{x}$.} that is increasing, differentiable with derivative $g$, and sigmoidal\footnote{A function is sigmoidal, if it is non-decreasing, concave to the right of a particular point called the inflection point and convex to its left. The second derivative of $\atana{x} = g'(x) = -\alpha x [1+x^2]^{-(1+\alpha/2)}$. The inflection point for $\atana{\cdot}$ is therefore 0.}. We may therefore write the received power at location $x_j$ (and therefore the interference in the single-frequency case) as
\begin{equation}
\label{eqn:Eo-altformula}
  E^o(x_j) = \atana{L-x_j} + \atana{L+x_j}.
\end{equation}
The following is a useful property of $E^o$.

\begin{proposition}
\label{prop:monotoneI}
$E^o$ is an even function with a unique maximum at $0$. Moreover, $E^o(|x|)$ monotonically decreases with $|x|$.
\end{proposition}

\begin{IEEEproof}
See Appendix \ref{app:proofMonotoneI}.
\end{IEEEproof}

\subsubsection{The two-frequencies case}
Unlike the previous setting, in which the two BSs operate on the same RF band, in the two-frequencies case the total interference at each BS depends on the association decisions of the mobiles. Indeed, the interference power at BS~$j$ is the total power received at that BS from all mobiles that actually associate with it. The total received power at  BS~$j$ is thus given by $E(x_j, A_j)$.

For example, suppose $A_1 := [-L ,\theta] $ and $A_2 := (\theta , L ]$ denote the two \emph{cells} for some $\theta \in [-L,L]$. Then the interference power at BS~1 is $E(x_1, A_1) = \atana{\theta-x_1} - \atana{-L-x_1}$. The expression for $E(x_2,A_2)$ is obtained analogously.

\subsection{SINR-equilibrium association}
We shall first consider the case in which the BSs' locations are fixed, and each mobile has the option of associating with one of the BSs. \emph{The continuum of mobiles constitute the players in this association game.}

Consider a mobile at location $y$. \emph{Its utility is a nondecreasing function of the throughput density at $y$}~(see Appendix \ref{app:prop-fluid}). The throughput density at $y$ increases linearly with SINR density. Thus, this mobile chooses a BS that yields the higher SINR density at $y$. Let $I_j$ be the set of interferers as seen at BS~$j$. If a mobile at point $y$ is associated with BS~$j$, the SINR density for this mobile is 
\begin{equation}
  \SINR(y,x_j,I_j) := \frac{g(y - x_j)}{E(x_j,I_j) + \sigma^2 }.  \label{eqn:sinr_expression}
\end{equation}
A mobile at $y \in [-L,L]$ will therefore prefer to associate with BS~1 if $\SINR(y,x_1,I_1) \geq \SINR(y,x_2,I_2)$.

We observe that in the single-frequency case, $I_j = [-L,L]$. Thus, the SINR density for a location, as seen at BS~$j$ is fixed. However, in the two-frequencies case, $I_j = A_j, j = 1,2$. Hence, the SINR density for a location, as seen at BS~$j$ is a function of the cell $A_j$.

\begin{definition}
  \label{def:sinr_equilibrium}
 The cell partition $(A_1,A_2)$ is said to be an SINR-equilibrium if the following holds: $y \in A_1$ if $\SINR(y,x_1,I_1) > \SINR(y,x_2,I_2)$ and only if $\SINR(y,x_1,I_1) \geq \SINR(y,x_2,I_2)$. If $\SINR(y,x_1,I_1) = \SINR(y,x_2,I_2)$, $y \in A_1$ or $A_2$ arbitrarily.
 \end{definition}

\begin{remark}
  This definition of equilibrium is similar to the Wardrop equilibrium in road traffic~\cite{wardrop}, or the Nash equilibrium in population games~\cite{Sandholm}. Note, however, that in Wardrop equilibrium the utility of choosing a resource~(a BS in the present problem) depends on the set of users that make the same choice through their total ``number'' (their fraction or their mass). Extensions of the Wardrop concept exist to the case where there is a finite number of user classes and the utility of using a resource for a user in a given class depends on the amount of users of each one of the classes who use that resource~\cite{dafermos}. In our problem, however, there is a continuum of classes corresponding to the locations of the mobiles. 
\end{remark}

\subsection{Hierarchical equilibrium}
We shall also consider placement of BSs taking into account the SINR-equilibrium that follows when mobiles associate to maximize their SINR density. \emph{The two BSs play a location game:} BS~$j$ decides to place itself at $(x_j,1)$ where $x_j \in \mathbb{R}, j = 1,2$. \emph{The utility of a BS is a monotone function of the aggregate throughput of all the mobiles associated with it}. Since the throughput density at location $y$ increases linearly with SINR density, we may simply set the integral of SINR density over the cell of a BS as its utility. Thus for BS~$j$ with cell $A_j$ and interferers  $I_j$, the utility is
\[
  \frac{1}{2} \int_{A_j}{\SINR(y,x_j,I_j)}~dy  = \frac{1}{2} \int_{A_j}{\frac{g(y - x_j) ~dy}{E(x_j,I_j) + \sigma^2 }}
\]
Once the BSs choose their locations, $A_j$, $I_j$, and thus the utility of  BS~$j$ are determined by the association game played by the mobiles. We thus have a Stackelberg-like game~\cite{CC} with the lead players being the two BSs (who may either cooperate or compete) and the followers the continuum of mobiles (who compete to maximize their respective SINR densities). We refer to this as the hierarchical equilibrium problem.

\section{CDMA: The Single-Frequency Case}

\subsection{SINR-equilibrium association}
\label{subsec:sinrEqAssoc}
We begin by providing closed form expressions for cell
boundaries in the SINR-equilibrium (see Definition~\ref{def:sinr_equilibrium}). 
Define the $\alpha$th root of the ratio of the net interferences (including thermal
noise) at the two BSs to be
\[
  B_{\alpha}(x_1, x_2) := \left( \frac{E^o(x_1) + \sigma^2} {E^o(x_2) + \sigma^2} \right)^{1/\alpha}.
\]
We start by considering symmetric placements of BSs: $|x_1| = |x_2|$ which implies $B_{\alpha}(x_1, x_2)=1$.
If $x_1 = x_2$ BSs are indifferent to all the mobiles from the point of view of SINR density.
Hence $(A,[-L,L]\backslash A)$ for all $A \subset [-L,L]$ are SINR-equilibrium association profiles. 
If $x_1 = -x_2 \neq 0$, mobiles associate with the BS which is closer. Hence either $([-L,0],(0,L])$~(if $x_1 < 0$)
or $([0,L]),[-L,0))$~(if $x_1 > 0$) is the unique SINR-equilibrium.
To study the asymmetric scenarios we assume, without loss of generality, that BS~2 is
located closer to the origin than BS~1, i.e., $|x_1| > |x_2| \geq 0$.
On account of Proposition~\ref{prop:monotoneI}, we have $B_{\alpha}(x_1,x_2) < 1$.

\begin{proposition}
  \label{prop:cellBoundaries}
  Let BS~1 be located at $x_1$ and BS~2 at $x_2$ where $|x_1| > |x_2| \geq
  0$. The set of mobile locations that connect to BS~2 is nonempty 
  only if
  \begin{equation}
    \label{eqn:nonEmptyCondition}
    \tau := |x_2 - x_1| \cdot \frac{B_{\alpha}(x_1, x_2)}{1-B_{\alpha}^2(x_1, x_2)} \geq 1.
  \end{equation}
  If the inequality holds strictly then the set of locations that connect to
  BS~2 is given by the interval\footnote{The notation $a+(b,c)$ is short for the interval $(a+b, a+c)$.}
  \[
  \frac{x_2 - x_1 B_{\alpha}^2(x_1, x_2)}{1 - B_{\alpha}^2(x_1, x_2)} + \left( - \sqrt{\tau^2 -1} ,  \sqrt{\tau^2 - 1} \right).
  \]
\end{proposition}


\begin{IEEEproof}
   Mobiles that have a higher SINR density at $x_2$ will connect to BS~2, i.e., $y \in A_2$ if
\[
  \frac{[(y - x_2)^2 + 1]^{-\alpha/2}}{E^o(x_2) + \sigma^2} > \frac{[(y - x_1)^2 + 1]^{-\alpha/2}}{E^o(x_1) + \sigma^2}
\]
which is equivalent to
\[
  (y - x_2)^2 + 1 < \left( (y - x_1)^2 + 1 \right) B_{\alpha}^2(x_1, x_2).
\]
As $B_{\alpha}^2(x_1,x_2) < 1$, the above inequality holds when a convex quadratic function of $y$ is strictly negative. The positive discriminant condition straightforwardly yields that the set connecting to BS~2 is nonempty only if~(\ref{eqn:nonEmptyCondition}) holds. The roots of the convex quadratic equation are given by the ends of the specified interval. Since the convex quadratic function is strictly negative in the interval between the roots, all the mobiles in this interval have higher SINR densities at BS~2.
\end{IEEEproof}

When $|x_2| > |x_1| \geq 0$, the roles of BS~1 and BS~2 are switched: BS~1 sees more interference, its cell $A_1$ may be empty, and when nonempty, $A_1$ is an interval.

\begin{figure}[h]
\centering
\includegraphics[width=3.49in, height=3.5in]{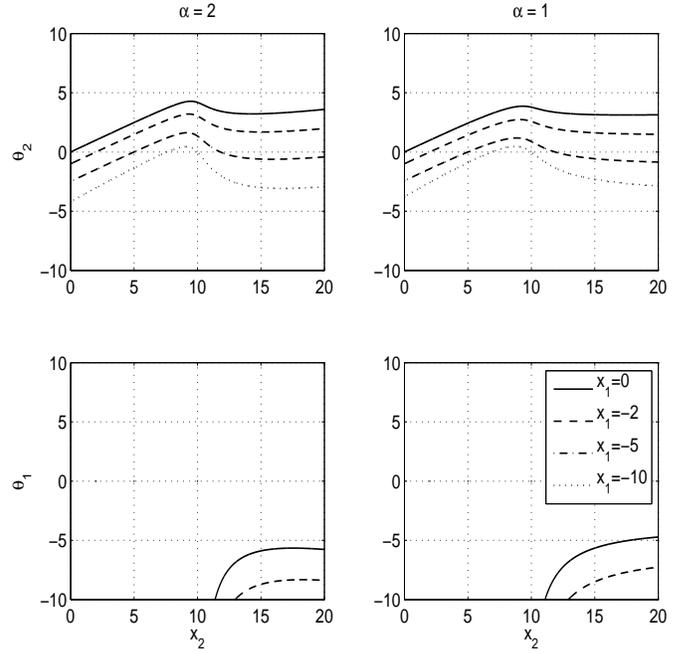}
\caption{Single-frequency case; SINR-equilibrium: Thresholds determining the cell boundaries as a function of the location of BS~2 for various locations of BS~1. The path loss exponent is 2 in the figures on the left and 1 in those on the right.}
\label{fig:thres}
\end{figure}

We provide numerical results to illustrate some surprising features of the SINR-equilibrium that distinguish this from other association games~(e.g.,~\cite{hotelling,dafermos}). We set $L=10$ (so that mobiles are concentrated over the interval $[-10,10]$) and the noise parameter $\sigma=0.3$. We place BS~1 at one of the fixed locations $x_1$ where $x_1= - 10, -5, -2, 0$. For each of these, we vary the location of BS~2 from $x_2=0$ to $x_2=30$~(see Figure~\ref{fig:thres}). The left column of plots corresponds to a path loss exponent $\alpha= 2$ and the right one to $\alpha=1$. The equilibrium sets $A_j$ turn out to have the form $A_1 = [\theta_1, \theta_2]$, $A_2 = [-L, \theta_1) \cup (\theta_2, L]$ for $x_1 = 0, -2$, and $A_1 = [-L, \theta_2]$, $A_2 = (\theta_2, L]$ for $x_1 = -5, -10$. The top~(respectively bottom) row of plots depict the threshold $\theta_2$~(respectively $\theta_1$) as a function of $x_2$. See the following for more details.

\subsubsection{Observations}
\paragraph{Non-convex cells}
For all the locations of BS~1,  $x_1= - 10, -5, -2, 0$, mobiles in $(\theta_2, L]$ have a  better SINR density at BS~2. Let us concentrate on the curves corresponding to $x_1= -2$ in Figure~\ref{fig:thres}. When BS~2 is located sufficiently far to the right of the origin, the interference at BS~1 is large compared to that at BS~2~(see Proposition~\ref{prop:monotoneI}). Thus, mobiles sufficiently far away and to the left of BS~1 (those in $[-L, \theta_1)$) also have a better SINR density at BS~2 despite BS~2 being the farther BS. Thus, in this case, $A_2 = [-L, \theta_1) \cup (\theta_2, L]$, a non-convex set. $A_2$ is similarly non-convex when $x_1= 0$ and $x_2$ is sufficiently far to the right (or left).

\paragraph{Non-monotonicity of the cell boundaries}

\begin{figure}[h]
\centering
\includegraphics[width = 2.5in, height = 2.2in]{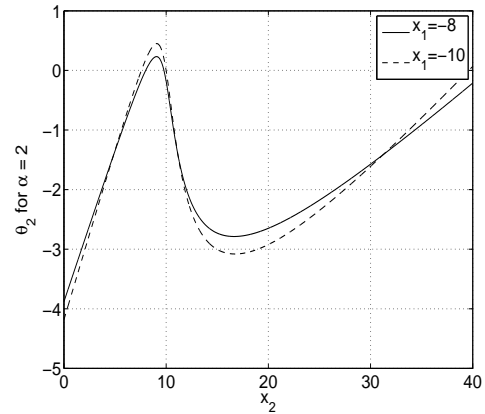}
\caption{Single-frequency case; SINR-equilibrium: Upper cell boundary of cell $A_1$ as a function of the location of BS~2 for various locations of BS~1.}
\label{fig:3thres}
\end{figure}

We observe a surprising non-monotonicity of the threshold $\theta_2$ as a function of the location $x_2$ of BS~2. $\theta_2$ first increases with $x_2$ until about $x_2=8$, then it decreases with $x_2$ until around $x_2=14$; finally, for larger $x_2$, $\theta_2$ again increases. Analogous non-monotonicity  is observed in $\theta_1$ too~(in the the curves corresponding to $x_1= -2,0$ and $\alpha = 2$).

The dashed line in Figure~\ref{fig:3thres} shows a zoomed-in view of the $x_1= -10$ case of the top-left plot of Figure~\ref{fig:thres}. The threshold $\theta_2$ increases beyond 0 until $x_2$ is about 8 units to the right of the origin, and then returns to 0 when $x_2=10$. This can be understood as follows. Clearly, for $x_2=10$ the interferences at both the BSs are the same; hence $\theta_2=0$, the midpoint. Now imagine moving BS~2 a little to the left (i.e., decreasing $x_2$). Now $|x_2| < |x_1|$. Thus, from Proposition~\ref{prop:monotoneI},  the interference $E^o(x_2)$ at BS~2 is larger than $E^o(x_1)$, the interference at BS~1. This makes it advantageous for mobiles a little to the right of the origin also to associate with BS~1; hence $\theta_2$ increases as $x_2$ decreases from $x_2=10$. Further decrease in $x_2$ brings BS~2 closer to mobiles on the negative $x$-axis, thus ultimately causing $\theta_2$ to return to 0, and even cross below 0, as $x_2$ decreases further. As $x_2$ increases beyond $10$, the interference perceived by it decreases, thus making it advantageous for mobiles a little to the left of the origin also to associate with BS~2; hence $\theta_2$ decreases as $x_2$ increases beyond $x_2=10$. Once BS~2 is moved far from the region where the mobiles exist, the signal power to $x_2$ becomes smaller, and association with BS~1 becomes increasingly better for mobiles to the right of the origin, causing $\theta_2$ to increase.

The top row of plots in Figure~\ref{fig:thres} suggests that $\theta_2$ is perhaps monotone in the position of BS~1. But this is not true because a closer look at the $\theta_2$ curves in Figure~\ref{fig:3thres} for $x_1=-10,-8$ shows that they cross each other several times.


\subsubsection{Discussion}
The form of equilibria displayed in the SINR-association examples is unusual in the class of location games. The reason for the unusual features lies in the SINR criterion, as we describe now.
\begin{itemize}
\item If a mobile is very close to a BS, path gain from the mobile to the BS will be very high. Thus, the mobile connects to this BS, even if the interference suffered by this BS is relatively higher.
\item If a mobile is located sufficiently far from both BSs, then the relative difference in the powers received at the BSs will be small. Thus the mobile will prefer to connect to BS that suffers from less interference.
\item If a mobile is at moderate distance from both the BSs, it takes into account both the factors (i) path gains to the BSs and (ii) interferences suffered by the BSs, while making its association decision.


\end{itemize}

\subsection{Hierarchical equilibrium}
\label{subsec:hierarchical}
\paragraph*{Single base station}
Suppose there is only one BS. Given that the interference is maximum at the origin and decreases monotonically with distance from the origin, where should it be placed to maximize utility? The utility of the BS, when placed at $x$, is given by
\begin{eqnarray*}
  \frac{1}{2} \int_{-L}^L \frac{g(y-x)}{E^o(x) + \sigma^2} ~dy = \frac{1}{2}\frac{E^o(x)}{E^o(x) + \sigma^2}
\end{eqnarray*}
which is maximized when $E^o(x)$ is maximized, i.e., at $x = 0$. Despite the high interference, the origin is the best location to maximize the utility given the nature of the utility function.

\paragraph*{Two base stations, utility behavior}

\begin{figure}[h]
\centering
\includegraphics[width=2.5in, height=2.2in]{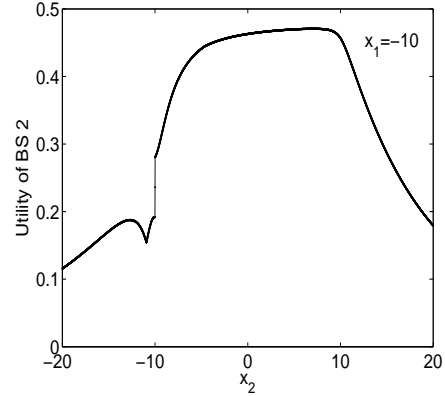}
\caption{Single-frequency case; Utility of BS~2 as a function of its location when we position BS~1 at $x_1 = -10$.}
\label{fig:fig12}
\end{figure}

Figure \ref{fig:fig12} portrays the effect of cell boundaries on utility. BS~1 is located at $x_1 = -10$. The utility of BS~2 as a function of its location is then plotted. For $x_2 \ll -10$, mobiles at the farthest right connect to BS~2. For $x_2$ approximately in the interval -11 to -10, an interval in the left most extreme of $[-10,10]$ also joins BS~2 so that BS~2 cell partition is a union of two intervals. Consequently, the utility increases in this interval. At $x_2 = -10$ a sudden transition occurs where all nodes in the middle interval switch to BS2, and hence the discontinuity in the utility. For $-10 < x_2 < 10$ nodes in an interval join BS~2, and this eventually becomes a half infinite line with boundary moving to the right as $x_2$ becomes large.

\subsubsection{Two cooperating base stations}
We now consider optimal joint placement of two BSs to maximize the sum utility. 


It can be shown that in a hierarchical optimal configuration the two BSs are placed on the opposite sides of the origin. 
Furthermore, the corresponding cells are of the form $[-L,a]$ and  $(a,L]$, and are 
characterized by a single parameter $a$. See Appendix~\ref{appen:sfc} for justifications. While the exact characterization of $a$ remains open, simulations indicate that sum utility is maximized when $a= 0$ and  $-x_1 = x_2$, i.e., the BSs are equidistant from the origin. 
We call such a placement as \emph{symmetric}. The SINR-equilibrium cells under symmetric placement are $[-L, 0]$ and $(0, L]$.

\begin{remark}
The optimal configuration
should also be an SINR-equilibrium association profile implying that for a user at $a$, 
SINR densities seen at BS~1 and BS~2 must match, i.e.,
\[
  \frac{g(a - x_1)}{E^o(x_1) + \sigma^2} = \frac{g(a - x_2)}{E^o(x_2) + \sigma^2}.
\]
Note that the symmetric configuration with boundary $a = 0$ and $-x_1 = x_2$ satisfies this condition.
\end{remark}

\begin{figure}[h]
\centering
\includegraphics[width=2.5in, height=2.2in]{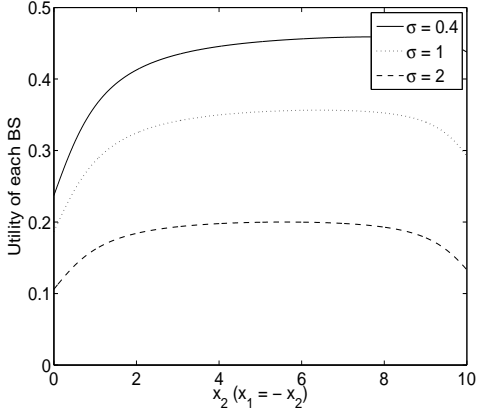}
\caption{Single-frequency case; cooperating BSs; symmetric placement of BSs at $\pm x$: Plot of utility obtained by each BS vs.\ x; here $L=10,\alpha=2$ and $\sigma$ takes the values 0.4, 1, and 2.}
\label{fig:stack1}
\end{figure}

Figure \ref{fig:stack1} depicts the utility obtained by each BS for symmetric placement $-x_1 = x_2 = x$, as a function of $x$. We see that  the origin and the extreme points (at distance 10 from the origin) are suboptimal locations. We also observe that the performance close to the optimal location is quite robust to perturbations of BS locations.

\begin{figure}[h]
\centering
\includegraphics[width=2.5in,height=2.2in]{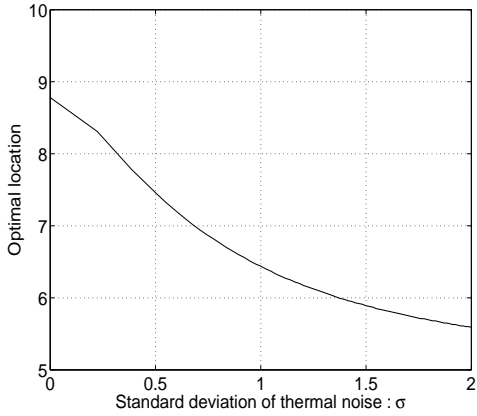}
\caption{Single-frequency case; cooperating BSs; symmetric placement of BSs: Optimal symmetric distance from origin for two BSs, as a function of thermal noise standard deviation.}
\label{fig:optimalSymmetricPlacement}
\end{figure}

Further experimentation revealed that, as $\sigma$ is increased, optimal distance of the BSs from the origin decreases (see Figure~\ref{fig:optimalSymmetricPlacement}). As $\sigma \rightarrow \infty$, the optimal symmetric locations of the BSs converge to -5 and 5. This is expected because at very large $\sigma$, interference does not play any role, and the BSs should be placed to maximize the total power collected from the respective cells, $E(x,(0,L]) + E(-x, [-L,0])$. Proposition~\ref{prop:monotoneI} says that this is maximized by choosing $x$ and $-x$ to be the mid-points of the respective intervals, i.e., $x = L/2$, which is 5 in our example.

\subsubsection{Two non-cooperating base stations}
\label{nonC}
We now consider a non-cooperative game between the two BSs. 
The BSs act simultaneously and pick their locations to maximize their respective utilities.


\begin{figure}[h]
\centering
\includegraphics[width=2.5in, height=2.2in]{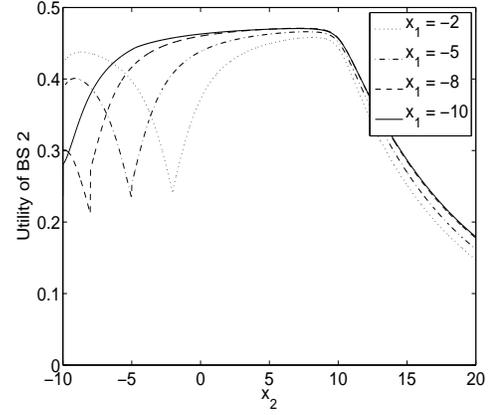}
\caption{Single-frequency case; non-cooperating BSs: Utility of BS~2 as a function of its location
when we position BS~1 at $x_1$ where $x_1=-2,-5,-8,-10$.}
\label{fig:stk-ass2}
\end{figure}
Figure \ref{fig:stk-ass2} has on the horizontal axis the location of BS~2 and on the vertical axis the utility it achieves. The figure is obtained for $L=10, \sigma=0.3, \alpha=2$. There are four curves that correspond to four locations of BS~1: $x_1= -2,-5,-8,-10$. From these curves, one can conclude that the utility of BS~2 is quite robust to placement errors around the best response location, for the indicated values of BS~1 locations.

\begin{figure}[h]
\centering
\includegraphics[width=2.5in,height=2.2in]{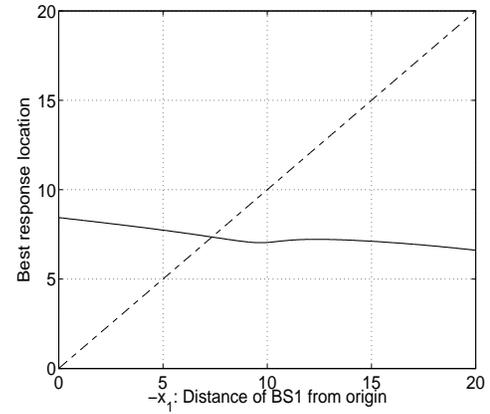}
\caption{Single-frequency case; non-cooperating BSs: The best response of BS~2 when BS~1 is at a distance indicated by abscissa to the left of the origin.}
\label{fig:nonCoopBestResponse}
\end{figure}
Figure \ref{fig:nonCoopBestResponse} shows the best response of BS~2 to a BS~1 location. BS~1 is moved along the segment to left of the origin. In the figure the horizontal axis is $-x_1$, distance of BS~1 from the origin. A positive best response value indicates a location on the other side of the origin away from BS~1. Numerical computations indicate the existence of a unique symmetric equilibrium at $-x_1 = x_2 = 7.36$.

In Table \ref{table1} we compare the optimal location of the cooperative case and the equilibrium location of the non-cooperative case, as a function of $\sigma$. We observe that at the non-cooperative equilibrium, the BSs are closer than at the cooperative optimum, i.e., placements are more aggressive. An analytical proof of this observation remains an interesting open problem. In both cooperative and non-cooperative cases the distances decrease in $\sigma$ and tend to a limit which is $-x_1=x_2=5$ for the cooperative case and $-x_1 = x_2 = 4.06$ for the non-cooperative case. 

\begin{table}[h]
\caption{Optimal cooperative and non-cooperative placements of BSs as a function of
$\sigma$} 
\label{table1}
\centering{
\begin{tabular}{|c|ccccc|} \hline
$\sigma$  & 0.1 & 0.4 & 1 & 2 & 40
\\
\hline Optimum distance  & & &  & & \\
of BSs from 0 & 8.658 &  7.745 & 6.435 & 5.591 & 5.002 \\
(cooperative) & & & & & \\
\hline Equilibrium distance & & & & & \\
of BSs from 0 & 8.10 &  6.95 & 5.50 & 4.667 & 4.09
\\ (non-cooperative) & & & & & \\
\hline
\end{tabular}
}
\end{table}

\section{CDMA: The Two-Frequencies Case}
\label{sec:twoFrequencies}

\subsection{SINR-equilibrium association}
\label{subsec:sinrEqAssoc-2freq}
We study the properties of the SINR-equilibrium partition and arrive at a numerical method to compute it.
The case where BSs are collocated, i.e., $x_1 = x_2$, is trivial. Let us assume that $x_1 \neq x_2$.
Recall that the interference at BS~$j$ in the two-frequencies case is $E(x_j, A_j)$. As in Section~\ref{subsec:sinrEqAssoc}, 
define $B$ to be the $\alpha$th root of the ratio of the net interferences at the two BSs, i.e.,
\[
  B := \left[ \frac{E(x_1,A_1) + \sigma^2}{E(x_2,A_2) + \sigma^2} \right]^{1/\alpha}.
\]
Note that $B \in [B_{\min}, B_{\max}]$, where 
\[B_{\min} =  \left[ \frac{\sigma^2}{E^o(x_2) + \sigma^2} \right]^{1/\alpha} \mbox{and} \ B_{\max} =  \left[ \frac{E^o(x_1) + \sigma^2}{\sigma^2} \right]^{1/\alpha}.\]
A location $y \in A_2$ if 
\begin{eqnarray}
  \lefteqn{ \frac{g(y-x_2)}{E(x_2,A_2) + \sigma^2} >  \frac{g(y-x_1)}{E(x_1,A_1) + \sigma^2}, } \nonumber \\
  & \Longleftrightarrow & (y-x_2)^2 + 1 < \left( (y-x_1)^2 + 1 \right) B^2. \label{eqn:cond-y}
\end{eqnarray}
First consider $B < 1$. Proceeding exactly as in the proof of Proposition~\ref{prop:cellBoundaries}, we get $A_2$ to be the set of $y$ such that a convex quadratic function of $y$ is negative. Thus $A_2$ is an interval and its complement $A_1$ a union of at most two intervals. More precisely, the boundaries are given as follows.  Define
\begin{equation*}
    \tau(B) := |x_1 - x_2| \cdot \left| \frac{B}{1-B^2} \right|.
\end{equation*}
If $\tau(B) \leq 1$, $A_2$ is empty, $B^2 = (E(x_1,A_1)+\sigma^2)/\sigma^2 > 1$, a contradiction. Thus $\tau(B) > 1$ and $A_2$ is determined by the interval (as in the proof of Proposition~\ref{prop:cellBoundaries})
  \begin{eqnarray*}
  \lefteqn{ (g_1(B), g_2(B))  } \\
  & :=  &  \frac{x_2 - x_1 B^2}{1 - B^2} + \left( - \sqrt{\tau(B)^2 -1} ,  \sqrt{\tau(B)^2 - 1} \right).
  \end{eqnarray*}
This gives expressions for the end points of intervals that make up $A_1$ and $A_2$ in terms of $B$. In particular,\footnote{$\lceil m \rceil_L = \min \{ m, L \}$, $\lfloor m \rfloor_{-L} = \max \{ m, -L \}$, and $[m]_{-L}^L = \min \{ \max \{ m, -L \}, L \}$}
\begin{eqnarray*}
A_1 &=& [-L, \lfloor g_1(B) \rfloor_{-L}) \cup (\lceil g_2(B) \rceil_L, L], \\
A_2  &=& [\lfloor g_1(B) \rfloor_{-L}, \lceil g_2(B) \rceil_L]. 
\end{eqnarray*}
Similar analysis could be done for $B > 1$. In this case, it can be shown that 
\begin{eqnarray*}
A_1  &=& [\lfloor g_1(B) \rfloor_{-L}, \lceil g_2(B) \rceil_L],\\
A_2 &=& [-L, \lfloor g_1(B) \rfloor_{-L}) \cup (\lceil g_2(B) \rceil_L, L], 
\end{eqnarray*}
where $g_1(B),g_2(B)$, and $\tau(B)$ are the same as defined above. Finally for $B = 1$, (\ref{eqn:cond-y}) implies 
\[A_1 = \left[-L, \left[\frac{x_1 + x_2}{2}\right]_{-L}^L\right] \ \mbox{and} \ A_2 = \left[\left[\frac{x_1 + x_2}{2}\right]_{-L}^L, L\right].\]
To emphasize that $A_1$ and $A_2$ depend only on $B$, we write $A_1(B)$ and $A_2(B)$. At SINR-equilibrium, therefore, $B$ must be a solution to the fixed point equation
\begin{equation}
\label{eqn:fixedpointB}
  B  =  \left[\frac{E(x_1, A_1(B)) + \sigma^2}{E(x_2, A_2(B)) + \sigma^2} \right]^{1/\alpha}=: F(B).
\end{equation}
\begin{thm}
\label{fp-unique}
The fixed point equation $F(B) = B$ has a unique solution. 
\end{thm}
\begin{IEEEproof}
We first prove that $E(x_1, A_1(B))$ and $E(x_2, A_2(B))$ are continuous in $B$. By inspection, $g_1(B)$ and $g_2(B)$ are continuous 
for all $B \neq 1$. Straightforward calculations show that $g_1(B) \rightarrow (x_1 + x_2)/2$ and $g_2(B) \rightarrow \infty$ as $B \uparrow 1$, while $g_1(B) \rightarrow -\infty$ and $g_2(B) \rightarrow (x_1 + x_2)/2$ as $B \downarrow 1$.
 So the boundaries of $A_1(B)$ and $A_2(B)$, after restriction to $[-L,L]$, are continuous in $B$ in $[B_{\min},B_{\max}]$. Thus  $E(x_1, A_1(B))$ and $E(x_2, A_2(B))$~(see~(\ref{eqn:E-x-S})) and therefore $F(B)$ are continuous functions of $B$ in $[B_{\min},B_{\max}]$. 

Next we show that $F(B)$ is a decreasing function of $B$. Let $B \leq 1$. We observe that if a $y$ satisfies~(\ref{eqn:cond-y}) for some value of $B$, it will do so also for any larger value of $B$ in $[B_{\min},1]$. Thus, $A_2(B)$ is an increasing set function~(order is specified by inclusion relation), and $A_1(B)$ is a decreasing set function. For $B > 1$, a similar argument shows that with $B' = 1/B$, $A_1(B')$ is increasing in $B'$, and $A_2(B')$ is decreasing in $B'$. So $A_1(B)$ and $A_2(B)$ are decreasing and increasing set functions respectively, for $B$ in $[1,B_{\max}]$ as well. Obviously $E(\cdot,A)$ strictly increases as $A \subset [-L,L]$ increases. Hence $F(B)$ is a decreasing function of $B$ for $B$ in $[B_{\min},B_{\max}]$. 

Finally, from~(\ref{eqn:fixedpointB}), we see that $F(B) \in [B_{\min},B_{\max}]$. Thus there is a unique fixed point of the equation $F(B) = B$.  
\end{IEEEproof}

As a simple example, consider the symmetric case when $-x_1 = x_2 \neq 0$. It is easy to verify that $B =1$, and the unique equilibrium partition is $([-L,0], (0,L])$ if $x_1 < 0$ and $([0,L], [-L,0))$ if $x_1 > 0$. 

For an integer $\alpha \geq 1$,~(\ref{eqn:fixedpointB}) can be written as an implicit equation in $B$ as
\begin{eqnarray*}
  \label{eqn:twoFreqImplicitEqn}
 B^{\alpha} = \ \ \ \ \ \ \ \ \ \ \ \ \ \ \ \ \ \ \ \ \ \ \ \ \ \ \ \ \ \ \ \ \ \ \ \ \ \ \ \ \ \ \ \ \ \ \ \ \ \ \ \ \ \ \ \ \ \ \ \ \ \ \ \ \\
  \frac{
          \begin{array}{l}
             \left[ \arctan_{\alpha}(L - x_1) - \arctan_{\alpha}(\lceil g_2(B) \rceil_L - x_1) \right. \\
             \left. + \arctan_{\alpha}(\lfloor g_1(B) \rfloor_{-L} - x_1) - \arctan_{\alpha}(-L - x_1)  + \sigma^2 \right]
          \end{array}
          }
     {\arctan_{\alpha}(\lceil g_2(B) \rceil_L - x_2) - \arctan_{\alpha}(\lfloor g_1(B) \rfloor_{-L} - x_2) + \sigma^2}. \nonumber
\end{eqnarray*}
We may numerically search for a $B$ that solves the above equation through a suitably fine quantization of the specified interval. 

\begin{remark}
Suppose $B^{\ast}$ solves the fixed point equation~(\ref{eqn:fixedpointB}). 
$B^{\ast} < 1$ implies that, at equilibrium, BS~2 has more interference than BS~1, and thus $A_2$ is a connected subset of $[-L, L]$.  $B^{\ast} > 1$ implies that, at equilibrium, BS~1 has more interference than BS~2, and thus $A_1$ is a connected subset of $[-L,L]$. 
\end{remark}
\begin{figure}[h]
\centering
\includegraphics[width=2.5in, height=2.2in]{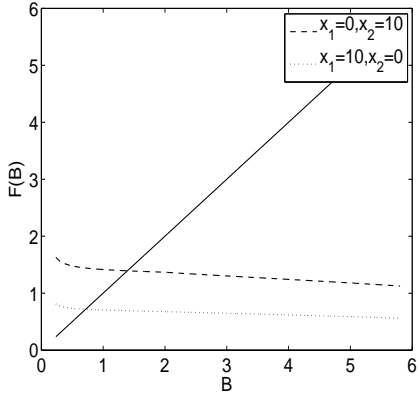}
\caption{Two-frequencies case; F(B) vs B. Figure illustrates the existence and uniqueness of fixed 
points.}
\label{fig:fbvsb}
\end{figure}
Figure~\ref{fig:fbvsb} illustrates the existence and uniqueness of fixed points. 
We set $L=10$, the path loss exponent $\alpha =2$ and the noise parameter $\sigma=0.3$.
Two curves are shown: one corresponding to 
$(x_1,x_2) = (0,10)$ and the other corresponding to $(x_1,x_2) = (10,0)$. 
For both the cases $F(B)$ is monotonically decreasing and cuts the line $y = x$ 
at unique points. In the first case, the fixed point is $B = 1.393$: at equilibrium, BS~1 has more interference than BS~2.
In the diagonally opposite second case, the fixed point $B = 0.718~( = 1/1.393)$. 

We also provide an algorithm to obtain the fixed point. Define $d := |x_1 - x_2|/2, \beta_{\min} := \sqrt{d^2+1} - d$ and $\beta_{\max} := \sqrt{d^2+1} + d$. Then, for $B < 1$, $\tau(B)>1$ implies $B \in (\beta_{\min}, 1]$. Similarly, for $B > 1$, $\tau(B)>1$ implies that $B \in [1, \beta_{\max})$. From an earlier discussion, the fixed point lies in $(\beta_{\min},\beta_{\max})$. It can also be verified that $F'(B) \rightarrow -\infty$ as $B \rightarrow \beta_{\min}$ or $\beta_{\max}$ and is finite otherwise. Thus a fixed point iteration of the form $B_{n+1} = F(B_n)$~(see~(\ref{eqn:fixedpointB})) may not converge. Motivated by~\cite{KAMG}, we propose a variant of this iteration which always converges to the desired fixed point. See Appendix~\ref{appen:relaxed-fp} for the algorithm and its analysis.

\begin{figure}[h]
\centering
\includegraphics[width=3.49in, height=1.5in]{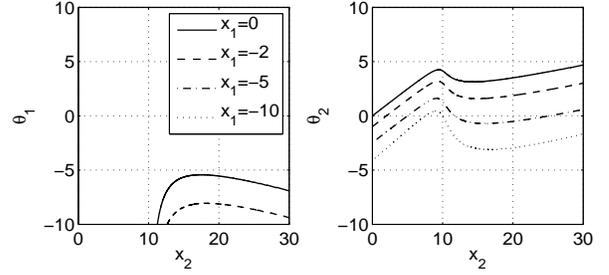}
\caption{Two frequencies case; SINR-equilibrium: Thresholds determining the cell boundaries as a function of the location of BS~2 for various locations of BS~1.}
\label{fig:thres2freq}
\end{figure}

We illustrate the form of SINR-equilibrium via some numerical results. We set $L=10$, the path loss exponent $\alpha =2$ and the noise parameter $\sigma=0.3$. We place BS~1 at one of the fixed locations $x_1$ where $x_1= - 10, -5, -2, 0$. For each of these, we vary the location of BS~2 from $x_2=0$ to $x_2=30$~(see Figure~\ref{fig:thres2freq}). As in single frequency case, the equilibrium sets $A_j$ have the form $A_1 = [\theta_1, \theta_2]$, $A_2 = [-L, \theta_1) \cup (\theta_2, L]$ for $x_1 = 0, -2$, and $A_1 = [-L, \theta_2]$, $A_2 = (\theta_2, L]$ for $x_1 = -5, -10$. The left~(respectively right) plot depicts the threshold $\theta_1$~(respectively $\theta_2$) as a function of $x_2$. 

\subsection{Hierarchical equilibrium}

\subsubsection{Two cooperating base stations}
The goal here is to place the two BSs so that the sum utility is maximized.

\begin{proposition}
\label{prop:2freq-coop}
  The locations $-x_1 = x_2 = L/2$ with SINR-equilibrium cell partition $(A_1, A_2) = (~[-L, 0], (0, L] ~)$ maximizes the sum utility.
\end{proposition}

\begin{IEEEproof}
For a given pair of locations $x_1$ and $x_2$, let $(A_1, A_2)$ be the SINR-equilibrium cell partition. For convenience let $u_j := E(x_j, A_j), ~j=1,2$, be the received power at BS~$j$. Then the sum utility satisfies the following:
\begin{eqnarray}
  \sum_{j=1}^2 \frac{1}{2} \frac{u_j}{u_j + \sigma^2} & \leq &  \frac{(u_1+u_2)/2}{(u_1+u_2)/2 + \sigma^2} \label{eqn:2freq-coop-SchurUB}\\
  & \leq & \frac{u_{\max}/2}{u_{\max}/2 + \sigma^2} \label{eqn:2freq-coop-monotone}\\
  & = & \frac{E(L/2, [0,L])}{E(L/2, [0,L]) + \sigma^2} \label{eqn:2freq-coop-max}
\end{eqnarray}
where (\ref{eqn:2freq-coop-SchurUB}) follows from Jensen's inequality because the function $u/(u+\sigma^2)$ is concave in $u$; inequality (\ref{eqn:2freq-coop-monotone}) follows because the function $\frac{u/2}{u/2 + \sigma^2}$ is monotone increasing in $u$ with $u_{\max}$ the maximum sum of received energies across any partition (not just SINR-equilibrium partitions). The last equality (\ref{eqn:2freq-coop-max}) follows from Proposition~\ref{prop:maxSumEnergy} in Appendix~\ref{app:SumEnergy}. The upper bound is independent of $x_1$ and $x_2$, and is achieved when $-x_1 = x_2 = L/2$. The corresponding intervals indeed constitute an SINR-equilibrium cell partition.
\end{IEEEproof}

\subsubsection{Two non-cooperating base stations}
\label{twofeq:nonC}
We now consider the hierarchical game where the BSs compete with each other keeping in mind their individual utilities as in Section \ref{nonC}.

\begin{figure}[h]
\centering
\includegraphics[width=2.5in, height=2.2in]{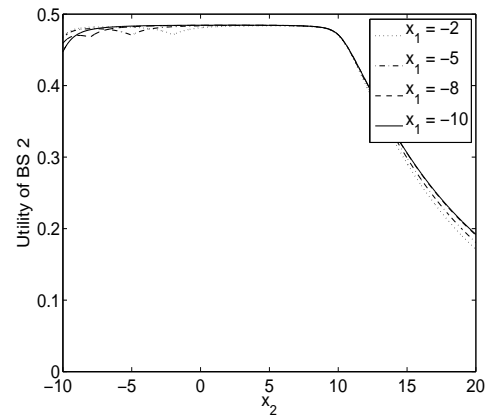}
\caption{Two frequencies case; non-cooperating BSs: Utility of BS~2 as a function of its location
when we position BS~1 at $x_1$ where $x_1=-2,-5,-8,-10$.}
\label{fig:stk-ass2-2freq}
\end{figure}

\begin{figure}[h]
\centering
\includegraphics[width=2.5in,height=2.2in]{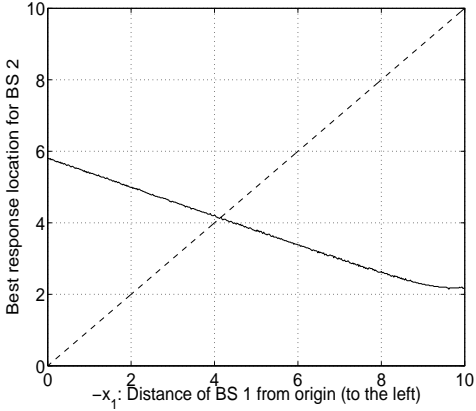}
\caption{Two frequencies case; non-cooperating BSs: The best response of BS~2 when BS~1 at a distance indicated by abscissa to the left of the origin. A positive best response indicates a location on the other side of BS~1.}
\label{fig:gameDiffFre2BS}
\end{figure}

Figure \ref{fig:stk-ass2-2freq} has on the horizontal axis the location $x_2$ of BS~2 and on the vertical axis the utility it achieves. The figure is obtained for $L=10, \sigma=0.3, \alpha=2$. There are four curves that correspond to four locations of BS~1: $x_1= -2,-5,-8,-10$. From these curves, one can conclude that the utility of BS~2 is quite robust to placement errors around the best response location, for the indicated values of BS~1 locations.
Figure \ref{fig:gameDiffFre2BS} yields the best response for BS~2 given BS~1's placement. Given a BS~1 location, the higher interference cell and the equilibrium ratio $B$ are first found as discussed in Section \ref{subsec:sinrEqAssoc-2freq}, for each possible location of BS~2. Then the BS~2 location yielding the maximum utility is identified as the best response location and is plotted in the figure.

Numerical results indicate that there is a unique symmetric equilibrium for the chosen parameters at $-x_1 =  x_2 = 4.1 $. The corresponding SINR-equilibrium cell partition is $A_1 = [-L, 0]$ and $A_2 = (0,L]$. Note that any unilateral deviation will change the cell boundaries and yield lesser utility to the deviating BS. If the BSs were cooperative, the best locations are $-x_1 = x_2 = 5$. However, this latter set of locations is not an equilibrium under competition.
We also performed numerical computations for other values of noise variance, e.g., $\sigma = 0.1,1,2$. It is observed that the equilibrium BS locations are insensitive to the value of $\sigma$. Yet again, we observe  in our numerical examples that placements are more aggressive~(i.e., BSs are closer to each other) under competition.

\section{SIC: The Single-Frequency Case}
\label{sec:sicsinglefreq}
We now extend our study to incorporate the effect of employing SIC decoding by the BSs. 
For model description, see Appendix \ref{app:prop-fluid}. In this section, we consider the single frequency case where mobiles connected to BS~$1$ cause interference at BS~$2$ and vice-versa.

\subsection{SINR-equilibrium association}
We first study the properties of the SINR-equilibrium partition.
The SINR density seen at BS~$j$, for a mobile at $y \in A_j$, can range from $\frac{g(y-x_j)}{E^o(x_j) + \sigma^2}$ to $\frac{g(y-x_j)}{E^o(x_j) - E(x_j, A_j) + \sigma^2}$, depending on the BS's decoding order. We assume that each mobile first associates with a BS. The BSs then choose an arbitrary decoding order. In the absence of a clear policy for choosing the decoding order at the BSs, we assume that the mobiles at $y$ associate to BS~$2$ only if 
\[\frac{g(y-x_2)}{E(x_2,A_1) + \sigma^2} \geq  \frac{g(y-x_1)}{E(x_1,A_2) + \sigma^2}.\]
We may interpret this as an association where a mobile {\em optimistically} believes that it will be decoded last~(at the BS with which it is associated) and therefore expects to see an SINR density $\frac{g(y-x_j)}{(E^o(x_j) - E(x_j, A_j) + \sigma^2)}$ with BS~$j$. 
Without loss of generality, relabel indices so that
\[
  B := \left[ \frac{E(x_1,A_2) + \sigma^2}{E(x_2,A_1) + \sigma^2} \right]^{1/\alpha} \leq 1.
\]
The above condition can be rewritten as 
\begin{equation*}
  (y-x_2)^2 + 1 \leq \left( (y-x_1)^2 + 1 \right) B^2,
\end{equation*}
where $B^2 \leq 1$. 

The condition governing the structure of the SINR-equilibrium partition has the same form as the one in Section~\ref{subsec:sinrEqAssoc}. 
As before, the cell partition $(A_1, A_2) = ([-L,0], (0,L])$ is an SINR-equilibrium with $B = 1$ in the symmetric case when $-x_1 = x_2$. 

Numerical computations show that there can be more than one SINR-equilibrium partitions for a pair of BS locations~$(x_1,x_2)$.
To illustrate this phenomenon, assume that all the mobiles are associated with BS~$1$. A mobile's anticipated SINR densities at the two BSs would be $\frac{g(y-x_1)}{\sigma^2}$ and $\frac{g(y-x_2)}{E^o(x_2) + \sigma^2}$ respectively. It does not switch to BS~$2$ if
\begin{equation*}
  \frac{g(y-x_2)}{E^o(x_2) + \sigma^2} <  \frac{g(y-x_1)}{\sigma^2}. 
\end{equation*}
Thus all mobiles keep their associations if
\begin{equation*}
 \frac{g(y-x_2)}{g(y-x_1)} < 1 + \frac{E^o(x_2)}{\sigma^2}, \ \forall \ y \in [-L,L].
\end{equation*}
This makes $([-L,L],\emptyset)$ an SINR-equilibrium partition.
Similarly one can argue that $(\emptyset,[-L,L])$ is also an SINR-equilibrium partition if
\begin{equation*}
\frac{g(y-x_1)}{g(y-x_2)} < 1 + \frac{E^o(x_1)}{\sigma^2}, \ \forall \ y \in [-L,L].
\end{equation*}
If the noise variance $\sigma^2$ is sufficiently small, both conditions can hold and hence both $([-L,L],\emptyset)$ and $(\emptyset,[-L,L])$ are SINR-equilibrium partitions.

\begin{remark}
The above discussion illustrates  an interesting {\em capture} phenomenon, which is of interest if the two BSs are placed sequentially, i.e., a Stackelberg game is played. The BS being placed first can judiciously place itself, to capture a majority of the mobiles. 
\end{remark}

\subsection{Hierarchical equilibrium}
BS~1 and BS~2 choose their respective locations cooperatively. 
Each BS then employs SIC decoding for all mobiles in its cell.
From the discussion in Appendix \ref{app:prop-fluid}, the utility of BS~$1$ is $\frac{1}{2} \log \left( 1 + \frac{E(x_1, A_1)}{E(x_1, A_2) + \sigma^2} \right)$, independent of the decoding order. A similar expression is obtained for the utility of BS~$2$.

\subsubsection*{Two cooperating base stations}
In the cooperative case, if $\sigma^2 \approx 0$, it is nearly optimal if all mobiles can associate to one BS. This is because if there is a non-zero population of mobiles connected to one BS, it generates a non-zero interference to the other BS. On the other hand, with all the mobiles associated to one of the BSs, say BS~1, the sum utility $\frac{1}{2} \log \left( 1 + \frac{E^o(x_1)}{\sigma^2} \right) \rightarrow \infty$ when $\sigma^2 \rightarrow 0$. So BS~2 should be placed very far away so that its cell is nearly empty. Symmetric placements are therefore not optimal in general. However, as $\sigma \rightarrow \infty$, interference from the other BS no longer plays a role. As in the single mobile decoding scenario, BSs should be placed to maximize the total power collected from their respective cells. Thus symmetric locations $-x_1 = x_2 = \frac{L}{2}$ become optimal.

Recall that there can be multiple SINR-equilibria for a given pair of BSs locations. Thus for the case of   
non-cooperating BSs, the competitive equilibria are not well defined. We do not pursue their study in this work.

 
\subsection{Pessimistically behaving mobiles}
\label{subsec:pessimistic-mobiles}
In the absence of a clear policy for choosing the decoding order at the BSs, mobiles might also think pessimistically, i.e., each mobile makes an association decision assuming that it will be decoded first~(at the BS with which it associates) and therefore expects to see an SINR density $\frac{g(y-x_j)}{(E^o(x_j) + \sigma^2)}$ with BS~$j$. This is the same SINR-density as observed in the case of single mobile decoding. Hence for given BS placements, SINR-equilibrium partition will be identical to that in the case of single mobile decoding~(Section~\ref{subsec:sinrEqAssoc}).

For cooperating BSs, hierarchical equilibrium  will be asymmetric in general as in the case of optimistic mobiles.
When BSs are selfish, BS~$1$ is interested in optimizing
\begin{eqnarray}
\lefteqn{\frac{1}{2} \log \left(1 + \frac{E(x_1, A_1)}{E(x_1, A_2) + \sigma^2} \right)} \nonumber \\
&=& \frac{1}{2} \log \left(1 + \frac{\frac{E(x_1, A_1)}{E^o(x_1) + \sigma^2}}{1 - \frac{E(x_1, A_1)}{E^o(x_1) + \sigma^2}} \right) \label{sic-cdma}.
\end{eqnarray}
 This is equivalent to optimizing $\frac{E(x_1, A_1)}{E^o(x_1) + \sigma^2}$ which is the utility of BS~$1$
if single mobile decoding is employed. In the single mobile decoding case numerical examples indicate that there is a hierarchical equilibrium with symmetric BS placements, say $-x_1 = x_1 = x^{\ast}$, and  cell partition $(A_1, A_2) = ([-L,0], (0,L])$~(Section~\ref{nonC}). Clearly, the same is a hierarchical equilibrium in the case of SIC decoding as well.  For a BS, the aggregate equilibrium utilities corresponding to single mobile decoding and SIC decoding, are related as in~(\ref{sic-cdma}).

 \begin{remark}
Having discussed the two extreme decoding order beliefs, we are naturally led to the following interesting problem. What decoding policy should a BS advertise in order to maximize its utility? Recall that given an association profile, a BS's utility does not depend on what decoding order it actually follows. But the advertisement will affect the SINR-equilibrium and thus the  utility. We leave this as an open problem for future research. 
\end{remark} 

\section{SIC: The Two-Frequencies Case}
\label{sec:sictwofreq}
We now proceed to the two-frequencies case where the BSs employ SIC decoding. We give a complete characterization of both cooperative and competitive equilibria. 

\subsection{SINR-equilibrium association}
In the two frequencies case, the SINR density seen at BS~$j$, for a mobile at $y \in A_j$, can range from $\frac{g(y-x_j)}{E(x_j, A_j) + \sigma^2}$ to $\frac{g(y-x_j)}{\sigma^2}$, depending on the BS's decoding order. As before, we assume that each mobile first associates with a BS. The BSs then choose an arbitrary decoding order. In the absence of a clear policy for choosing the decoding order at the BSs, an optimistic mobile believes that it will be decoded last and therefore expects to see an SINR density of $g(y-x_j)/\sigma^2$ with BS~$j$. This being monotonically decreasing in the distance $|y-x_j|$, the mobile simply associates to the nearest BS. If the BSs are collocated, then either BS is chosen arbitrarily. Define $v = (x_1 + x_2)/2$. Then, the equilibrium cell partition is $A_1 = [-L, v], ~A_2 = (v, L]$ if $x_1 < x_2$, $A_1 = [v, L], ~ A_2 = [-L, v)$ if $x_1 > x_2$, and an arbitrary choice at every $y$ if $x_1 = x_2$.

\subsection{Hierarchical equilibrium}
In the two frequencies case, BS~$j$'s utility with SIC decoding is $\frac{1}{2} \log (1 + \frac{E(x_j, A_j)}{\sigma^2}), ~j=1,2$~(see Appendix~\ref{app:prop-fluid}).
\subsubsection{Two cooperating base stations}
In this case, the two BSs cooperate to maximize sum utility. 
\begin{thm}
  Consider the two-frequencies case with two cooperating BSs that employ SIC decoding. The BS locations that maximize sum throughput are $-x_1 = x_2 = L/2$.
\end{thm}

\begin{IEEEproof}
Recall the notation used in the proof of Proposition~\ref{prop:2freq-coop} where $u_j = E(x_j,A_j)$. The sum throughput may be upper bounded as
\begin{eqnarray*}
  \sum_{j=1}^2 \frac{1}{2} \log \left( 1 + \frac{u_j}{\sigma^2} \right)
    & \leq & \log \left( 1 + \frac{u_1 + u_2}{2 \sigma^2} \right) \\
    & \leq & \log \left( 1 + \frac{u_{\max}}{\sigma^2} \right) \\
    & = & \log (1+ \frac{E(L/2, (0,L])}{\sigma^2}).
\end{eqnarray*}
where the first inequality follows from Jensen's inequality, while the second follows as in the proof of Proposition~\ref{prop:2freq-coop}, and the third follows from Proposition~\ref{prop:maxSumEnergy}. Finally, the upper bound is attained at $-x_1 = x_2 = L/2$ with cell partition $(~[-L,0], (0,L] ~)$. This completes the proof.
\end{IEEEproof}

Consider now a case where the BSs are constrained to be collocated at $x$. Recall that mobiles pick one or the other BS with equal probability, so that the power collected at each BS is $E^o(x)/2$ yielding a sum utility $\log \left(1 + \frac{E^o(x)}{2 \sigma^2} \right)$. This attains its maximum when $E^o$ does, which is at $x = 0$ (see Proposition~\ref{prop:monotoneI}).

\subsubsection{Two non-cooperating base stations}
In this case, the two BSs play a non-cooperative game to maximize their respective utilities. Define $a := 2^{2/\alpha}$. For $\alpha \in [1,\infty)$, we have $a \in (1,4]$. Recall that if the two BSs are not collocated, the cell boundary is $(x_1 + x_2)/2$. Let $r_j(x_1,x_2)$ be the power collected by BS~$j$. The utility of each BS is a monotone function of the power collected, and we may therefore assume that BS~$j$'s goal is to maximize $r_j(x_1,x_2)$. Let $\BR_2(x_1)$ be the set of best responses of BS~2 to $x_1$, i.e., 
\[\BR_2(x_1) = \argmax_{x_2} r_2(x_1,x_2)\] 
Define $\BR_1(x_2)$ analogously.
\begin{lma}
\label{lma:l1}
(i) For all $x_1$, $\BR_2(x_1) \subseteq [-L,L]$. An analogous conclusion holds for $\BR_1(x_2)$. \\
(ii) If $(x_1, x_2)$ is an equilibrium strategy profile, then $x_1,x_2 \in [-L,L]$. 
\end{lma}
\begin{IEEEproof}
We consider the following three cases
\begin{enumerate}
\item If $x_1 \in (-\infty,-L)$, then $r_2(x_1,-L) = E^o(L)$. On the other hand if $|x_2| > L$,
then $r_2(x_1,x_2) \leq E^o(x_2) < E^o(L)$. Thus if  
an $x_2 \notin [-L,L]$, then $x_2 \notin \BR_2(x_1)$.

\item If $x_1 \in [-L,L]$, then 
\begin{eqnarray*}
r_2(x_1,-L) &>& r_2(x_1,x_2), \ \forall x_2 \in (-\infty,-L),\\
\mbox{and} \ r_2(x_1,L) &>& r_2(x_1,x_2), \ \forall x_2 \in (L,\infty).
\end{eqnarray*}
Thus if an $x_2 \in (-\infty,-L) \cup (L,\infty)$, then  $x_2 \notin \BR_2(x_1)$.

\item If $x_1 \in (L,\infty)$, then $r_2(x_1,L) = E^o(L)$. On the other hand if $|x_2| > L$,
then $r_2(x_1,x_2) \leq E^o(x_2) < E^o(L)$. Thus if  
an $x_2 \notin [-L,L]$, then $x_2 \notin \BR_2(x_1)$.
\end{enumerate}
Similar arguments hold for $\BR_1(x_2)$ also.

The second part follows immediately once we recognize that   $(x_1, x_2)$ is an equilibrium strategy profile
if and only if $x_1 \in \BR_1(x_2)$ and $x_2 \in \BR_2(x_1)$.
\end{IEEEproof}

Thus, for equilibrium analysis, we only need to focus on $x_1,x_2 \in [-L,L]$. For $x_1,x_2 \in [-L,L]$, $r_2(x_1,x_2)$ is as given in Table \ref{table:r2}, with a similar table for $r_1(x_1,x_2)$ of BS~1. Interestingly, the function $r_2(x_1, \cdot)$ as a function of $x_2$ is discontinuous at $x_2 = x_1$ unless $x_1 = 0$. A similar observation holds for $r_1(\cdot,x_2)$. 
\begin{table}[h]
\caption{Power received at BS~2: $x_1 \in [-L,L]$}
\label{table:r2}
\centering{
    \begin{tabular}{c|c}
    \hline
   $x_2 \in$ & $r_2(x_1,x_2)$ \\
    \hline
     $[-L, x_1)$ & $\atana{\frac{x_1 - x_2}{2}} + \atana{L+x_2}$  \\
          $\{x_1\}$  & $E^o(x_1)/2$ \\
$(x_1, L]$ & $\atana{L-x_2} + \atana{\frac{x_2 - x_1}{2}}$  \\
    \hline
    \end{tabular}
}
\end{table}
Next we show that in an equilibrium, the two BSs are placed on the opposite sides 
of the origin. 
\begin{lma}
If $x_1 < 0$, then $\BR_2(x_1) > x_1$. Similarly, if $x_1 > 0$, then $\BR_2(x_1) < x_1$.
Analogous conclusions hold for $\BR_1(x_2)$ also.
\end{lma}
\begin{IEEEproof}
Let $x_1 < 0$. If $x_1 < -L$, the result follows from Lemma~\ref{lma:l1}. For $-L \leq x_1 < 0$
\begin{eqnarray*}
r_2(x_1,x_1^+) &>& r_2(x_1,x_1),\\
\mbox{and} \ r_2(x_1,-x_2) &>& r_2(x_1,x_2), \ \forall x_2 < x_1.
\end{eqnarray*}
See Table~\ref{table:r2} to verify the first claim above. The second one follows since 
mobiles associate to the nearest BS. These two imply that if $x_2 \leq x_1$, then $x_2 \notin \BR_2(x_1)$.
This is the desired result. 
Similarly one can argue for $x_1 > 0$, and subsequently for $\BR_1(x_2)$.
\end{IEEEproof}
\begin{cor}
\label{cor:6.2}
If $(x_1, x_2)$ is an equilibrium strategy profile, then either~(i) $x_1 \in [-L,0], x_2 \in [0,L]$, 
or~(ii) $x_1 \in [0,L], x_2 \in [-L,0]$. 
\end{cor}
\begin{IEEEproof}
If $x_1 = 0$, there is nothing to prove. First let $x_1 \in [-L,0)$. The above lemma implies that
$x_2 \nleq x_1$, because  $x_2 = \BR_2(x_1) > x_1$. Moreover, $x_2 \notin (x_1, 0)$, because $x_1 = \BR_1(x_2) > x_2$ for $x_2 < 0$. 
Thus $x_2 \in [0,L]$. Case~(ii) is similarly handled.
\end{IEEEproof}

We now characterize all the equilibria.
\begin{thm}
\label{Thm6point2}
(i) For $L \leq \sqrt{a-1}$, there exists a unique equilibrium at $x_1 = x_2 = 0$. \\
(ii) For $L > \sqrt{a-1}$, there exists a unique equilibrium~(up to a permutation) at $-x_1 = x_2 = \frac{1}{a-1} \left( -L + \sqrt{aL^2 - (a-1)^2} \right)$.
\end{thm}
\begin{IEEEproof}
See Appendix \ref{app:proofThm6point2}.
\end{IEEEproof}

\begin{remark}
(i) The equilibria locations do not depend on $\sigma$. A similar insensitivity observation was made for the equilibria in single mobile 
decoding case~(Section~\ref{twofeq:nonC}). \\
(ii) 
Note that 
\[\frac{ -L + \sqrt{aL^2 - (a-1)^2}}{a-1} < \frac{  -L + \sqrt{aL^2}}{a-1} = \frac{L}{\sqrt{a}+1} < \frac{L}{2}.\] 
 Again, as already seen in Sections \ref{nonC} and~\ref{twofeq:nonC} for the case of single mobile decoding, the competitive equilibrium locations of BSs are closer to each other than the optimal locations under cooperation.
\end{remark}

\begin{figure}[h]
\centering
\includegraphics[width=3.49in, height=1.7in]{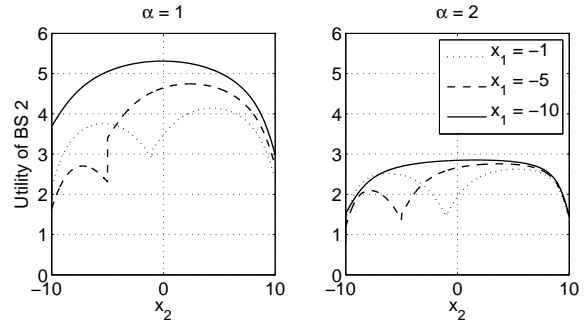}
\caption{SIC two frequencies case: $r_2(x_1,x_2)$ as a function of location $x_2$ of BS2 for three fixed locations  of BS1: $x_1= -1,-5,-10$. Here $L=10$ and $\alpha=1$ (left) and $\alpha=2$ (right).}
\label{sic}
\end{figure}

Figure \ref{sic} plots the received power at BS~2 given by $r_2(x_1,x_2)$ as a function of location $x_2$ for three fixed locations of BS~1: $x_1=-1,-5,-10$. Here $L=10$ and $\alpha=1$ (left) and $\alpha=2$ (right). Both satisfy $L \geq \sqrt{\alpha - 1}$. A clear best response location $x_2$ is seen for each fixed $x_1$. As BS~1 approaches the origin, the best response location of BS~2 moves further away from the origin.

\subsubsection{Convergence to equilibrium}
We consider the best response dynamics in which the location of each of the two BSs is sequentially adjusted.

\begin{thm}
\label{Thm6point3}
Let $L > \sqrt{a-1}$. Assume that BSs follow the best response dynamics to adjust their positions. Then, starting from arbitrary initial positions $x_1^o$ and $x_2^o$, the best response sequence converges to the unique equilibrium.
\end{thm}
\begin{IEEEproof}
See Appendix \ref{app:proofThm6point3}.
\end{IEEEproof}

\begin{figure}[h]
\centering
\includegraphics[width=2.5in, height=2.2in]{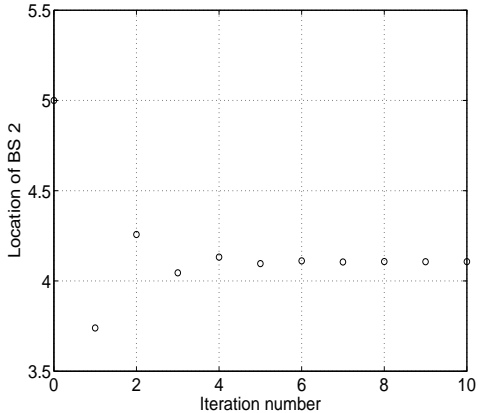}
\caption{SIC two frequencies case: non-cooperative BSs; Convergence of the best response dynamics for BS~2.}
\label{conv}
\end{figure}

As an example, consider $L= 10$ and $\alpha = 2$, i.e., $a = 2$.  
The equilibrium locations of the BSs are at a distance of 4.107 from the origin.
Figure \ref{conv} illustrates the fast convergence of dynamics from the starting locations $\pm 5$ to the equilibrium locations.

\begin{remark}
When mobiles behave pessimistically, i.e.,  each mobile makes its association decision assuming that it will be decoded first~(at the BS with which it associates), SINR-equilibrium and hierarchical equilibrium results are identical to those for the single mobile decoding in the two-frequencies case given in Section~\ref{sec:twoFrequencies}. (See the discussion in Section~\ref{subsec:pessimistic-mobiles} for a justification in the single-frequency case).
\end{remark}

\section{Conclusions}
\label{sec:conclusions}
We studied combined BS placements and mobile associations in a game-setting where the utilities were determined by SINR criteria. We saw that the SINR-equilibrium cells exhibited non-monotonicity and non-convexity properties that are not seen in the classical location game problems. These unusual properties arise because the SINR density that determines association is a function of the distance between a mobile and the BS it is associated with and also the BS location. We studied hierarchical equilibria in the CDMA single-frequency and two-frequencies cases. We saw evidence (via simulations in the CDMA single-frequency case and via analysis in the other case) of a unique optimal pair of locations in the cooperative scenario. We also saw evidence of a unique equilibrium pair of locations (up to permutation) in the competitive scenario. For the SIC single-frequency case, we made some interesting observations. For the SIC two-frequencies case, we completely characterized the optimal cooperative locations and all competitive equilibria. Interestingly, in all scenarios considered, the BS locations are closer to each other in the competitive case than in the cooperative case, an observation whose proof has eluded us.

\appendices

\section{Propagation, path loss, and fluid models}
\label{app:prop-fluid}
\paragraph*{Propagation model} A mobile transmitter is modeled as a point source that radiates in two-dimensional space or three-dimensional space. The wavefronts emanating from the point source are circular (respectively, spherical in three-dimensional space). We assume that the far field model holds and that antenna couplings between neighboring transmitters and between transmitters and receiver are negligible, even in the limit as mobiles get closer to each other.

\paragraph*{Path loss model} Under the far-field model for propagation in two dimensions with circular wavefronts, a receiver at a distance $r$ from the point source and having aperture arc width $s \ll r$ will capture only $s/(2\pi r)$ of the total transmitted power, so that propagation loss is proportional to $1/r$. If there is further dissipation in the medium (analogous to shadowing and scattering of electromagnetic waves in three dimensions) we model the propagation loss as proportional to $1/r^{\alpha}$, where $\alpha \geq 1$. The path loss model $1/r^{\alpha}$ for three dimensional propagation with $\alpha \geq 2$ is of course the standard one.

\paragraph*{Fluid model} Consider $n$ mobiles located on a line at positions $-L + j \Delta y + \frac{\Delta y}{2}, ~j=0,1,\cdots,n-1$ with separation spacing $\Delta y = \frac{2L}{n}$. We use the letter $y$ to represent the discrete location for finite $n$, and the continuum location $y \in (-L,L)$ when $n \rightarrow \infty$. Each mobile has power $\Delta p(y) = \Delta y$, so that we may think of transmitted power density per unit distance $dp/dy$ as 1 power unit per unit distance, and the total transmitted power as $2L$ power units. Consider the BS located at $x$ at a height of 1 unit from the line. The path loss for a mobile at $y$ is $g(y -x) = \left[ 1 + (y-x)^2 \right]^{-\alpha/2}$ (see~(\ref{eqn:g})). The total received power at the BS, if all of these are in the same frequency band, is
\begin{eqnarray*}
  E_n(x) & := & \sum_{y = -L + \Delta y / 2}^{L - \Delta y / 2} g(y - x) \Delta y  \\
   & \rightarrow & \int_{-L}^L g(y-x) dy ~ = ~ E^o(x),
\end{eqnarray*}
where the limit is taken as $n \rightarrow \infty$. Similarly, the total received power from mobiles in a set $A \subseteq [-L,L]$ is
\begin{eqnarray*}
  E_n(x,A) & = & \sum_{y \in A} g(y - x) \Delta y  \\
   & \rightarrow & \int_{A} g(y-x) dy ~ = ~ E(x,A).
\end{eqnarray*}
$E(x,A)$ was defined in (\ref{eqn:E-x-S}) and $E^o(x)$ was defined as $E(x,[-L,L])$ immediately after.

\paragraph*{SINR Density and throughput} Let $I \subseteq [-L,L]$ denote the set of locations that may be considered as interferer locations. The SINR is then
\[
  \mbox{SINR}_n(y,x,I) = \frac{g(y-x) \Delta y}{\sigma^2 + \sum_{y \in I} g(y - x) \Delta y}
\]
As $n \rightarrow \infty$, the denominator tends to $E(x,I) + \sigma^2$, the numerator goes to 0, and the ratio 
\[\frac{\mbox{SINR}_n(y,x,I)}{\Delta y} \rightarrow \frac{g(y-x)}{E(x,I) + \sigma^2},\] 
so that the latter may be thought of as SINR density (SINR per unit distance). Using Shannon's capacity formula for Gaussian channels, the data rate for a mobile at location $y$ is
\[
  \frac{1}{2} \log \left( 1 + \mbox{SINR}_n(y,x,I) \right) \approx \frac{1}{2} \mbox{SINR}_n(y,x,I)
\]
where the natural logarithm is employed and the unit of information is nats. (1 nat = $1/(\log 2)$ bits $\approx$ 1.44 bits). The aggregate throughput of mobiles in a set $A \subseteq [-L,L]$ is 
\begin{eqnarray*}
 \lefteqn{\sum_{y \in A} \frac{1}{2} \mbox{SINR}_n(y,x,I)}  \\
 &\rightarrow & \frac{1}{2} \int_A \frac{g(y-x)}{E(x,I) + \sigma^2} ~dy = \frac{1}{2}\frac{E(x,A)}{E(x,I) + \sigma^2},
\end{eqnarray*}
which is taken as the utility of a BS in the continuum case.

\paragraph*{SIC decoding} Let the interval $A \subseteq [-L,L]$ denote a set of locations associated with the BS at $x$. Suppose that the BS employs SIC. An arbitrary decoding order is chosen and communicated with the transmitters. For concreteness, let us assume that mobiles are decoded in the decreasing order of $y$ in $A$. Then all mobiles in $A$ that are to the left of a given user at $y$ will become interferers to $y$. The throughput for user at $y \in A$ is therefore
\begin{eqnarray*}
  \lefteqn{ \frac{1}{2} \log \left( 1 + \frac{g(y-x) \Delta y} {\sigma^2 + \sum_{y' < y, y' \in A} g(y'-x) \Delta y} \right) } \\
  & = & \frac{1}{2} \log \left( \sigma^2 + \sum_{y' \leq y, y' \in A} g(y'-x) \Delta y \right) \\
  && ~ - ~\frac{1}{2} \log \left( \sigma^2 + \sum_{y' < y, y' \in A} g(y'-x) \Delta y \right).
\end{eqnarray*}
Summing these up over discrete $y \in A$, and passing to the limit, we get the aggregate throughput of all the mobiles in set $A$ to be
\[\frac{1}{2} \log \left( 1 + \frac{\sum_{y \in A} g(y-x) \Delta y} {\sigma^2} \right) \rightarrow  \frac{1}{2} \log \left( 1 + \frac{E(x,A)}{\sigma^2} \right),\]
an expression that is used in Sections~\ref{sec:sicsinglefreq} and~\ref{sec:sictwofreq} for the utility of BS. Note that this remains the sum utility regardless of the decoding order chosen at the BS. Of course, the data rates for each mobile will depend on its position in the decoding order. The sender and the receiver should agree on this data rate and employ an appropriate code. 

\paragraph*{Discussion} It should be noted that the two-dimensional propagation (when $\alpha \in [1,2)$) and our treatment of mobiles as fluid particles on a line are merely caricatures of real life propagation models. The purpose of their study is to get a qualitative feel for what one might expect in the three-dimensional propagation model with mobiles distributed in a plane and receiver antennas placed at a height from the plane. See also the extension in Appendix~\ref{app:2D}.

\section{Proof of Proposition~\ref{prop:monotoneI}}
\label{app:proofMonotoneI}
That $E^o$ is an even function, is obvious from (\ref{eqn:Eo-altformula}). To see the monotonicity,
for $x \geq 0$, write
\begin{eqnarray}
\label{eqn:I0-Ix-1} E^o(0) -  E^o(-x)   \hspace*{-.1in} & = & \hspace*{-.1in} \int_{-L}^{L} g(y) ~dy - \int_{-L+x}^{L+x} g(y) ~dy \\
\nonumber                        \hspace*{-.1in} & = & \hspace*{-.1in} \int_{-L}^{-L+x} g(y) ~dy - \int_{L}^{L+x} g(y) ~dy \\
\label{eqn:I0-Ix-3}              \hspace*{-.1in} & = & \hspace*{-.1in} \int_0^x [ g(y-L) - g(y+L) ] ~dy
\end{eqnarray}
where (\ref{eqn:I0-Ix-1}) follows from (\ref{Int3}), and (\ref{eqn:I0-Ix-3}) via a change of variable $y-L \leftarrow y$ in the first integral and $y+L \leftarrow y$ in the second. The integrand in (\ref{eqn:I0-Ix-3}) is positive for $y \in [0,x]$. This proves the monotonicity.

\section{CDMA single-frequency case: structure of cooperative hierarchical equilibria}
\label{appen:sfc}
\begin{figure}[h]
\centering
\includegraphics[width=3.49in, height=1.2in]{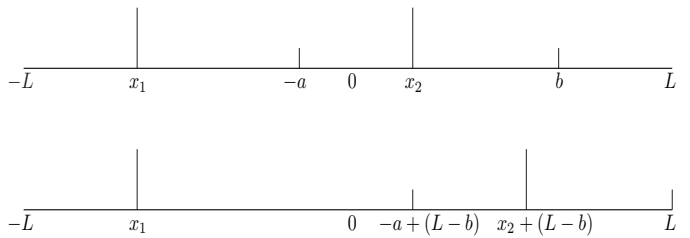}
\caption{In the above two figures $A_2 = [-a,b]$ and $[-a+(L-b),L]$, respectively. The top one can not be an optimal configuration; it is outperformed by the bottom one.}
\label{fig:fig21}
\end{figure}

\paragraph{In an optimal configuration the two BSs are placed on the opposite sides of the origin}
Assume that the two BSs are placed at $x_1,x_2$ such that $x_1,x_2 \leq 0$. 
The configuration $x_1 =x_2 < 0$ is outperformed by $x_1 = x_2 = 0$, and hence cannot be optimal. Next consider 
$x_1 < x_2 \leq 0$ without loss of generality. Under equilibrium, the cell $A_2$ of BS~2 is a convex set. Say $A_2 = [-a, b] \subseteq [-L,L]$. The following are observations that are easily seen.
\begin{itemize}
\item If $a < b$, the configuration $\{x_1,-x_2,A_2 = [-a,b],A_1 = [-L,L] \backslash [-a,b]\}$ has better joint performance.
\item If $a > b$, the configuration $\{x_1,-x_2,A_2 = [-b,a],A_1 = [-L,L] \backslash [-b,a]\}$ has better joint performance.
\item If $a = b$, the configuration $\{x_1,-x_2,A_2 = [-a,b],A_1 = [-L,L] \backslash [-a,b]\}$ performs equally well.
\end{itemize}  
In each case, the SINR-equilibrium of the modification can only have better joint performance.  This proves the claim.
\paragraph{In an optimal configuration, the cells of the two BSs are convex}
Without loss of generality, we may assume $x_1 \leq 0 \leq x_2$ and that $|x_1| \geq |x_2| \geq 0$.
Also assume that $A_2 = [-a,b]$ such that $a,b < L$ so that $A_1 = [-L,-a) \cup (b,L]$ 
is not convex~(see Figure~\ref{fig:fig21}). But then the configuration $\{ x_1,x_2+L-b,A_1 = [-L,-a+(L-b)),A_2=[-a+(L-b),L]\}$~(see Figure~\ref{fig:fig21}) strictly outperforms the assumed one.

Indeed, BS~2's utility remains the same, and a few of BS~1 mobiles have moved closer, keeping the 
same associated population size. This yields a contradiction and proves the convexity claim.

The above observation implies that in a hierarchical optimal configuration, the cells are of the form $[-L,a]$ and  $(a,L]$. 
Furthermore, the optimal configuration is also yielded by the following hierarchical problem.
For a given partition, i.e., a fixed $a$, the BSs are placed to optimize throughputs of the respective cells. However, the partition 
is made to maximize the sum utility of both the BSs while keeping in mind their subsequent selfish placements.

\section{The Relaxed Fixed Point Iteration}
\label{appen:relaxed-fp}
\paragraph*{The algorithm}
(i) First we determine the starting point. We aim at finding a point $B_0$ such that both $B_0$ and $F(B_0)$ lie in $(\beta_{\min},\beta_{\max})$. Recall that both $B$ and $F(B)$ belong to $[B_{\min}, B_{\max}]$ and $F(B)$ is decreasing in $B$. Also, we can easily verify that $F(B) = B_{\max}$ for all $B$ in\footnote{$[a,b] = \emptyset$ if $a > b$.} $[B_{\min},\beta_{\min}]$, and  $F(B) = B_{\min}$ for all $B$ in $[\beta_{\max},B_{\max}]$. A suitable choice of $B_0$ depends on the order of the quantities $B_{\min}, B_{\max}, \beta_{\min}$ and $\beta_{\max}$. The following are the four possible orders.~(Note that the fixed point lies in $(B_{\min}, B_{\max}) \cap (\beta_{\min},\beta_{\max})$, so the two intervals must intersect.) 
\begin{enumerate}
\item If $\beta_{\min} \leq B_{\min} < B_{\max} \leq \beta_{\max}$, choose a $B_0 \in (B_{\min}, B_{\max})$.
\item If $B_{\min} < \beta_{\min} < B_{\max} < \beta_{\max}$, there are two possibilities. 
\begin{enumerate}
\item If $F(B_{\max}) > \beta_{\min}$, choose $B_0 \in (\beta_{\min}, B_{\max})$.
\item If $F(B_{\max}) \leq \beta_{\min}$, choose $B_0 \in (\beta_{\min}, F^{-1}(\beta_{\min}))$.
\end{enumerate}
\item If $\beta_{\min} < B_{\min} < \beta_{\max} < B_{\max}$, again there are two possibilities.
\begin{enumerate}
\item If $F(B_{\min}) < \beta_{\max}$, choose $B_0 \in (B_{\min}, \beta_{\max})$.
\item If $F(B_{\min}) \geq \beta_{\max}$, choose $B_0 \in (F^{-1}(\beta_{\max}),\beta_{\max})$.
\end{enumerate}
\item If $B_{\min} < \beta_{\min} <  \beta_{\max} <  B_{\max}$, choose a $B_0 \in (F^{-1}(\beta_{\max}),$ $F^{-1}(\beta_{\min}))$.
\end{enumerate}
Set $\bar{B} = F(B_0)$. Our selection procedure ensures that $B_0$ and $\bar{B}$  are in $(\beta_{\min}, \beta_{\max})$. If $\bar{B} = B_0$, it is the desired fixed point. So, assume $\bar{B} > B_0$; the case $\bar{B} < B_0$ is handled similarly. From our earlier discussion $F'(B)$ is bounded over $[B_0, \bar{B}] \subsetneq (\beta_{\min}, \beta_{\max})$, i.e., there is a $D < \infty$ such that $|F'(B)| \leq D$ for all $B \in  [B_0, \bar{B}]$. Now, we focus only on this interval. \\
(ii) Choose a $\gamma \leq \frac{1}{1 + D}$, and define 
\begin{equation}
\label{eqn:iteration}
G(B) := \gamma F(B) + (1 - \gamma)B, \ B \in  [B_0, \bar{B}].
\end{equation}
(iii) Now iterate as $B_{n+1} = G(B_n), \ n = 0,1,\dots$.

\begin{proposition}
The iterates converge to the fixed point of $F(B) = B$. 
\end{proposition}
\begin{IEEEproof}
We prove that iterates as defined in Step~3) form a nondecreasing sequence. From~(\ref{eqn:iteration}), it suffices to show that $F(B_n) \geq B_n$ for all $n$. 
We show this inductively. 
Since $\bar{B} > B_0$, the claim is true for $n = 0$. Assume $F(B_n) \geq B_n$ for some $n$. From~(\ref{eqn:iteration}), we conclude that $B_{n+1} \geq B_n$.
Since $|F'(B)| \leq D$, we get 
\begin{equation}
\label{eqn:fp-inequality}
F(B_n) - F(B_{n+1}) \leq D(B_{n+1} - B_n).
\end{equation}
Moreover, our choice of $\gamma$~(see Step~(ii)) together with~(\ref{eqn:iteration}) ensures that
\[D(B_{n+1} - B_n) \leq F(B_n) - B_{n+1}.\]
Combining the above two inequalities we get $F(B_{n+1}) \geq B_{n+1}$. This completes the induction step.

Note that~(\ref{eqn:fp-inequality}) requires $B_n,B_{n+1} \leq \bar{B}$ which can also be shown inductively, as follows.
Since $B_1$ is a convex combination of $B_0$ and $\bar{B} := F(B_0)$, $B_1 \leq \bar{B}$.
Now assume $B_0 \leq B_1 \leq \dots B_n \leq \bar{B}$. This immediately implies that
\begin{eqnarray*}
B_{n+1} &=& \gamma F(B_n) + (1 - \gamma)B_n \\
        &\leq&  \gamma F(B_0) + (1 - \gamma)\bar{B} \\
        &=& \bar{B}.
 \end{eqnarray*}
Since the sequence $B_n, n = 0,1,\dots$ is bounded and nondecreasing, it converges. Finally, since $G(B)$ is a continuous function of 
$B$, the limit point $B^{\ast}$ is a fixed point of $G(B) = B$. Substituting in~(\ref{eqn:iteration}), we get $F(B^{\ast}) = B^{\ast}$. Therefore $B^{\ast}$ is the desired fixed point.
\end{IEEEproof}
\begin{figure}[h]
\centering
\includegraphics[width=3.49in, height=3.0in]{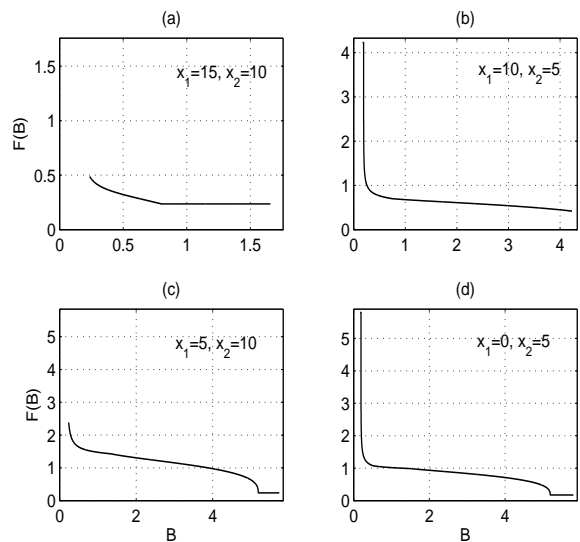}
\caption{Two-frequencies case; F(B) vs B. The four subplots illustrate the four possible scenarios as described in Step~$1$ of the relaxed fixed point iteration.}
\label{fig:rfpe}
\end{figure}
The four subplots in Figure~\ref{fig:rfpe} illustrate the four possible scenarios described in Step~$1$ above. We set $L = 10, \alpha = 2$ and $\sigma = 0.3$. Each subplot shows the corresponding locations of BSs. Note that, in all the cases, $d = |x_1 - x_2|/2 = 2.5$ which also fixes $(\beta_{\min},  \beta_{\max}) = (0.1926, 5.1926)$. $B_{\min}$ and $B_{\max}$ corresponding to the four subplots are as in Table~\ref{table2}.
\begin{table}[h]
\caption {$B_{\min}$ and $B_{\max}$  in the four subplots of Figure~\ref{fig:rfpe}}
\label{table2}
 \centering{
\begin{tabular}{|c|ccccc|} \hline
  & $x_1 = 15$ & $x_1 = 10$ & $x_1 = 5$ & $x_1 = 0$ & \\
 & $x_2 = 10$ & $x_2 = 5$ & $x_2 = 10$ & $x_2 = 5$ &\\
\hline 
$B_{\min}$ &  0.2364 & 0.1741 & 0.2364 & 0.1741 &\\
\hline 
$B_{\max}$ &  1.6580 & 4.2306 & 5.7423 & 5.8045 &\\
\hline 
\end{tabular}}
\end{table}

\begin{figure}[h]
\centering
\includegraphics[width=2.5in, height=2.2in]{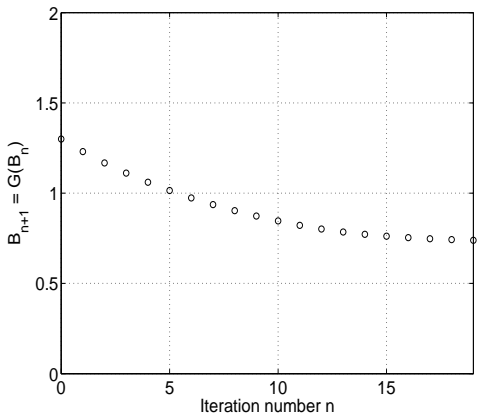}
\caption{Two frequencies case; convergence of the relaxed fixed point iteration.}
\label{fig:conv}
\end{figure}
Figure~\ref{fig:conv} shows convergence of the relaxed fixed point iteration via an example: $L=10, \alpha =2$  and $\sigma=0.3$; BS~1 and BS~2 locations are set to be $-20$ and $-15$ respectively. Thus $d = 2.5$, and $(\beta_{\min},  \beta_{\max}) = (0.1926, 5.1926)$.
Also $B_{\min} =0.6031$ while $B_{\max} =1.3180$. We set $\beta = 0.1$ and start the iteration with $B_0 = 1.3$. Iterations quickly converge to the unique fixed point $0.726$. 
 
\section{Two-frequencies case: sum received power}
\label{app:SumEnergy}
\begin{proposition}
  \label{prop:maxSumEnergy}
  Let $x_1 \leq x_2$ and let $v = (x_1+x_2)/2$ denote the mid-point. Then following results hold. \\
  (i) Let $(A_1, A_2)$ denote a partition of $[-L,L]$. Then 
  \begin{eqnarray*} 
     \lefteqn{ \max_{(A_1, A_2)} \left[ E(x_1, A_1) + E(x_2, A_2) \right] } \\
     && \leq E(x_1, [-L,v]) + E(x_2, (v,L]).
  \end{eqnarray*}
  (ii) Furthermore, $E(x_1, [-L,v]) + E(x_2, (v,L])$ is maximized at $-x_1 = x_2 = L/2$. The cell partition in this case is $[-L,0]$ and $(0,L]$.
\end{proposition}
\begin{IEEEproof}
  The first statement is obvious once we write out the integrals and recognize that the integrand is non-negative, symmetric, and $g(y)$ is decreasing in $|y|$. For the same reason, we may upper bound the sum in the second statement, by $E(x_1, I_1) + E(x_2, I_2)$ where $I_1$ is an interval of length $L+v$ centered at $x_1$ and $I_2$ is an interval of length $L-v$ centered at $x_2$. ($(I_1, I_2)$ may not be a partition of $[-L,L]$). Constraining the sum of interval lengths to be $2L$, the upper bound is further maximized when the intervals are of equal length $L$. But this upper bound is achieved when $-x_1 = x_2 = L/2$. The corresponding intervals $[-L,0]$ and $(0,L]$ constitute a cell partition, as required. This concludes the proof.
\end{IEEEproof}

\section{Proof of Theorem~\ref{Thm6point2}}
\label{app:proofThm6point2}
Let $(x_1,x_2)$ be an equilibrium. We may assume, without loss of generality, that $x_1 \in [-L,0], x_2 \in [0,L]$~(see Corollary~\ref{cor:6.2}). 
Differentiating $r_2(x_1,x_2)$ with respect to $x_2$, we get
\begin{equation}
\label{eqn:step1-1}
  \frac{\partial r_2(x_1,x_2)}{\partial x_2} = \frac{1}{2} g\left(\frac{x_2 - x_1}{2} \right) - g(L-x_2) 
\end{equation}
The rest of the proof is divided into two parts.
\paragraph*{Part~1} 
We first check whether $x_1 = x_2 = 0$ is an equilibrium. Note that $r_2(0,x_2)$ is an even, continuous and differentiable 
function of $x_2$.
From~(\ref{eqn:step1-1}) 
\begin{eqnarray*}
&& \frac{\partial r_2(0,x_2)}{\partial x_2}  = 0 \\
& \Longleftrightarrow & x_2^2\left(1 - \frac{a}{4}\right) - 2Lx_2 + L^2 - (a-1) = 0,
\end{eqnarray*}
which implies
\begin{eqnarray*}
x_2 = 
\left\{
\begin{array}{ll}
\frac{L \pm \sqrt{\frac{aL^2}{4} + (a-1)(1-\frac{a}{4})}}{1-\frac{a}{4}}, & a \in (1,4) \\
\frac{L^2 - (a-1)}{2L}, & a = 4
\end{array}
\right.
\end{eqnarray*}
Now, it is easy to check that $\frac{\partial r_2(0,x_2)}{\partial x_2} < 0$ for all $x_2 \in (0,L]$, if $L \leq \sqrt{a-1}$.
Thus, $\BR_2(0) = 0$. Analogously, $\BR_1(0) = 0$.
So, $(0,0)$ is an equilibrium, provided  $L \leq \sqrt{a-1}$. 
For $L > \sqrt{a-1}$,
\begin{eqnarray*}
\left. \frac{\partial r_2(0,x_2)}{\partial x_2 }\right|_{x_2 = 0^+} &=& \frac{1}{2}g(0) - g(L) \\
         &=& \frac{1}{2} - \frac{1}{(1 + L^2)^{\alpha/2}}\\
         &>& 0.
\end{eqnarray*}
Hence $\BR_2(0) \neq 0$. Thus $(0,0)$ can not be an equilibrium.
\paragraph*{Part~2} 
Next, let  $(x_1,x_2) \neq (0,0)$. Then, by assumption, we have $x_2 > x_1$.
From~(\ref{eqn:step1-1}), a best response $x_2$ to BS~1's location $x_1$ should satisfy
\begin{equation}
\label{eqn:step1-2}
  (L-x_2)^2 + 1 = a \left[ 1 + \left( \frac{x_2 - x_1} {2} \right)^2 \right].
\end{equation}
Similarly, a best response $x_1$ to BS~2's location $x_2$ should satisfy
\begin{equation}
\label{eqn:step1-3}
  (L+x_1)^2 + 1 = a \left[ 1 + \left( \frac{x_2 - x_1} {2} \right)^2 \right].
\end{equation}
Combining~(\ref{eqn:step1-2}) and~(\ref{eqn:step1-3}) we get
\[(x_1 + x_2)(x_2 - x_1 - 2L) = 0\]
If $x_2 - x_1 = 2L$, the only feasible candidate is $(x_1,x_2) = (-L,L)$. But, this contradicts~(\ref{eqn:step1-3}). Hence $x_2 = - x_1$. 
Again, from~(\ref{eqn:step1-3}), we have
\[ (L+x_1)^2 + 1 = a(1 + x_1^2) \]
which implies
\[x_1 = \frac{L \pm \sqrt{aL^2 - (a-1)^2}}{a-1}.\]
Since $x_1 \in [-L,0]$, we must have $\sqrt{aL^2 - (a-1)^2} > L$, so there will be no feasible solution for $L < \sqrt{a-1}$. Furthermore, 
\[x_1 = -x_2 = \frac{1}{a-1} \left( -L + \sqrt{aL^2 - (a-1)^2} \right)\]
 is the unique feasible candidate for $L \geq \sqrt{a-1}$. Finally, let $x^{\ast} := \frac{1}{a-1} \left( -L + \sqrt{aL^2 - (a-1)^2} \right)$. We now show that $x_1 = -x_2 = x^{\ast}$ is indeed an equilibrium if $L \geq \sqrt{a-1}$. 
To see this, note that
\[ \left. \frac{\partial^2 r_2(x^{\ast},x_2)}{\partial x_2^2}\right|_{x_2 = -x^{\ast}} = \frac{1}{4}g'(-x^{\ast}) + g'(L + x^{\ast}) <0,\] 
because $x^{\ast} < 0$  and $L + x^{\ast} > 0$. This assures that $- x^{\ast}$ is indeed a maximum point, i.e.,  $ - x^{\ast} = \BR_2(x^{\ast})$. Similarly, one can verify that  $x^{\ast} = \BR_1(-x^{\ast})$. 

Parts~1 and~2 together complete the proof. 

\section{Proof of Theorem~\ref{Thm6point3}}
\label{app:proofThm6point3}
Lemma~\ref{lma:l1} implies that it is sufficient to show the result for the case when, $x_1^o,x_2^o \in [-L,L]$.
Assume $x_1 \leq 0$. Solving~(\ref{eqn:step1-2}), we get
\begin{eqnarray}
\label{eq:thm5-2}
x_2 = 
\left\{
\begin{array}{ll}
\frac{4L - ax_1 - 2\sqrt{a(L- x_1)^2 + (4-a)(a-1)}}{4-a}, & a \in (1,4)\\
\frac{L + x_1}{2} - \frac{3}{2(L - x_1)}, & a = 4 
\end{array}
\right.
\end{eqnarray}
For $a \in (1,4)$, the solution to~(\ref{eqn:step1-2}) with the positive sign leads to $x_2 > L$.
We therefore discard it, because $\BR_2(x_1) \in [0, L]$. 
The other candidate is the $x_2$ in~(\ref{eq:thm5-2}) which satisfies $x_1 < x_2 < L$. This further implies that
 \[\frac{\partial^2 r_2(x_1,x_2)}{\partial x_2^2} = \frac{1}{4}g'\left(\frac{x_2 - x_1}{2}\right) + g'(L - x_2) <0. \] 
Hence $x_2$ as given in~(\ref{eq:thm5-2}) is indeed $\BR_2(x_1)$. 
Now, it can also be seen that $0 < \frac{\partial \BR_2(x_1)}{\partial x_1}  < 1 -\epsilon$ where $\epsilon>0$ depends on $a$. 
An analogous result holds for $x_1 > 0$ also. Analogous results also hold for  $\frac{\partial \BR_1(x_2)}{\partial x_2}$. Thus a small change in the position of a BS causes an even smaller change (in the same direction) in the position of the other BS. The best responses thus constitute a contraction map and the dynamics converges to the equilibrium.

\section{CDMA single-frequency case: Extension to Two Dimensions}
\label{app:2D}
A large number of mobiles are located uniformly over the two dimensional plain. Two BSs are placed at $(x_1,y_1,1)$ and $(x_2,y_2,1)$ respectively. The BSs operate on the same frequency band. The path loss model is as before. The SINR-equilibrium association can be defined in a similar way as in Definition~\ref{def:sinr_equilibrium}. Here, we provide closed form expressions for cell boundaries in the equilibrium. 
As in Section~\ref{subsec:sinrEqAssoc}, we define
\[
B_{\alpha} = \left( \frac{E^o(x_1,y_1) + \sigma^2} {E^o(x_2,y_2) + \sigma^2} \right)^{1/\alpha}.
\]
When the two BSs are collocated, every association profile is an SINR-equilibrium. 
If $(x_1,y_1) \neq (x_2,y_2)$ and still $B_{\alpha} = 1$, there is unique SINR-equilibrium association in which 
mobiles associate with the BS which is closer.
To study the asymmetric scenarios, without loss of generality assume that $B_{\alpha} < 1$, i.e., 
the interference at BS~2, $E^o(x_2,y_2)$, is more than that at BS~1, $E^o(x_1,y_1)$. 
Mobiles at $(x,y)$ connect to BS~2 if they do not have a lower SINR density at $(x_2,y_2)$, i.e.,
\begin{eqnarray*}
\lefteqn{  \frac{[(x - x_2)^2 + (y - y_2)^2 + 1]^{-\alpha/2}}{E^o(x_2,y_2) + \sigma^2} }\\
&\geq& \frac{[(x - x_1)^2 + (y - y_1)^2 + 1]^{-\alpha/2}}{E^o(x_1,y_1) + \sigma^2}
\end{eqnarray*}
which is equivalent to
\[
(x - x_2)^2 + (y-y_2)^2 + 1 \leq \left( (x-x_1)^2 + (y - y_1)^2 + 1 \right) B_{\alpha}^2.
\]
This inequality straightforwardly yields that the set connecting to BS~2 is nonempty only if
\[
\tau := \frac{\sqrt{(x_1-x_2)^2 + (y_1 - y_2)^2}B_{\alpha}}{1-B_{\alpha}^2} \geq 1.
\]
Furthermore, in case of a strict inequality above, the set is a disc with center
\[
\left(\frac{x_2 - x_1 B_{\alpha}^2}{1 - B_{\alpha}^2}, \frac{y_2 - y_1 B_{\alpha}^2}{1 - B_{\alpha}^2}\right) 
\]
and radius $\sqrt{\tau^2 -1}$.

Note that in the one-dimensional case, the BS with higher interference had an interval as its cell. The complement of this interval joined the second BS. The analog of the interval-type cell in two dimensions is a disc-type cell. Further, observe that we did  not make any assumptions on the population density. The cell partition is always a disc and its complement. However, the population density does affect the interference and therefore the actual parameters of the cell partition. 



\bibliographystyle{IEEEtran}
\bibliography{IEEEabrv,sinrgames.bib}

\end{document}